\documentclass[a4paper,fleqn,usenatbib]{mnras}

\usepackage{mathptmx}

\usepackage[T1]{fontenc}
\usepackage{ae,aecompl}

\usepackage{graphicx}	
\usepackage{amsmath}	
\usepackage{amssymb}	

\title[Suppression of star formation by quasar-driven winds?]{Evidence of suppression of star formation by quasar-driven winds in gas-rich host galaxies at z<1?}

\author[D.Wylezalek et al.]{
Dominika Wylezalek,$^{1}$\thanks{E-mail: dwylezalek@jhu.edu}
Nadia L. Zakamska,$^{2, 1}$
\\
$^{1}$Department of Physics \& Astronomy, Johns Hopkins University, Bloomberg Center, 3400 N. Charles St., Baltimore, MD 21218, USA\\
$^{2}$Deborah Lunder and Alan Ezekowitz Founders' Circle Member, Institute for Advanced Study, Einstein Dr., Princeton, NJ 08540, USA\\
}

\date{Accepted XXX. Received YYY; in original form ZZZ}

\pubyear{2016}

\begin{document}
\label{firstpage}
\pagerange{\pageref{firstpage}--\pageref{lastpage}}
\maketitle

\begin{abstract}
Feedback from active galactic nuclei (AGN) is widely considered to be the main driver in regulating the growth of massive galaxies through heating or driving gas out of the galaxy, preventing further increase in stellar mass. Observational proof for this scenario has, however, been scarce. We have assembled a sample of 132 radio-quiet type-2 and red AGN at $0.1<z<1$. We measure the kinematics of the AGN-ionized gas, the host galaxies' stellar masses and star formation rates and investigate the relationships between AGN luminosities, specific star formation rates (sSFR) and outflow strengths $W_{90}$ -- the 90\% velocity width of the [OIII]$\lambda$5007\AA\ line power and a proxy for the AGN-driven outflow speed. Outflow strength is independent of sSFR for AGN selected on their mid-IR luminosity, in agreement with previous work demonstrating that star formation is not sufficient to produce the observed ionized gas outflows which have to be powered by AGN activity. More importantly, we find a negative correlation between $W_{90}$ and sSFR in the AGN hosts with the highest SFRs, i.e., with the highest gas content, where presumably the coupling of the AGN-driven wind to the gas is strongest. This implies that AGN with strong outflow signatures are hosted in galaxies that are more `quenched' than galaxies with weaker outflow signatures. Despite the galaxies' high SFRs, we demonstrate that the outflows are not star-formation driven but indeed due to AGN-powering. This observation is consistent with the AGN having a net suppression, `negative' impact, through feedback on the galaxies' star formation history. \end{abstract}




\section{Introduction}

In the past decades, galaxy formation models have made significant progress at reproducing observed galaxy properties such as galaxy colors and the galaxy mass function \citep[e.g.][]{Croton_2006, Hopkins_2006, Somerville_2008, Novak_2011, Genel_2014, Alatalo_2014, Choi_2015, Smethurst_2015, Remus_2016}. These models invoke finding solutions for the so-called overcooling problem, where gas clouds cool too quickly and star formation happens too early \citep{Silk_2013}. If the gas had cooled at such a high rate as initially predicted by simulations, modern day galaxies would be much more massive than observed. Some sort of heating of gas is needed to prevent it from forming stars or gas removal from the host. Supernova explosions inject energy into the interstellar medium, but this energy release mechanism is insufficient to explain the observed low masses of nearby ellipticals \citep{Silk_2013}. 

The black hole at the center of a galaxy, however, can have a major impact on regulating the final stellar mass of the galaxy bulge \citep[e.g.][and references therein]{Fabian_2012}. The energy released during the growth of a black hole typically exceeds the binding energy of a galaxy by a large factor, such that even if only a few percent of that energy -- or even a smaller fraction \citep{AnglesAlcazar_2016} -- couples to the gas, the active galactic nucleus (AGN) can have a significant effect on the evolution of its host galaxy. AGN feedback and star formation quenching can explain the shape of the galaxy luminosity function and the bimodality in the distribution of galaxy colors. If AGN feedback couples to the gas as efficiently as required by galaxy formation models, then massive, early-type galaxies will be gas-poor and passively aging, while younger galaxies, typically lower-mass, late-type galaxies will still have enough star formation material and appear much bluer. 

Generally, two modes of such AGN feedback have been identified. During the \textit{kinetic mode}, also called radio jet or maintenance mode, the AGN luminosity is low and the feedback process is non-destructive. This mode keeps the balance between gas cooling and heating and is typically found in low redshift, massive galaxies. In the \textit{quasar mode} (also called radiative or wind mode) the AGN accretes close to the Eddington limit and AGN luminosities are high. The main interaction with the gas is through winds, and such energetic winds can deposit large amounts of kinetic energy into the interstellar medium and can drive high-velocity outflows. These outflows can have mass outflow rates of several 1000~M$_{\odot}$~yr$^{-1}$ \citep{Liu_2013b, Veilleux_2013, Cicone_2014, Brusa_2015a} which can deplete the galaxy of their gas on timescales of tens of megayears, resulting in the quenching of star formation. 

Many detailed studies of the gas fields in quasars (AGN with luminosities L$_{\rm bol}>10^{45}$~erg/sec) have shown that a high fraction of such galaxies indeed shows signatures of outflows with high velocities that cannot be reached through gas motion in a the gravitational potential of their host galaxies \citep{Liu_2013a, Liu_2013b, Harrison_2012}. These gas velocities are so high that only a powerful AGN could drive those \citep{Rupke_2013, Hill_2014}. Using the increasingly common technique of integral field unit spectroscopy and by thus exploring the velocity fields, spatial extents of both the gas from warm ionized interstellar medium and molecular gas, multiple groups have demonstrated that galaxy-wide winds extending up to several kilo-parsecs from the galaxy center seem to be common in powerful quasars both at high and lower redshift \citep{McElroy_2015, Brusa_2015a, Harrison_2016}.

While AGN feedback is now widely accepted to be the main driver in regulating the growth of galaxies, especially after the discovery of the tight relation between masses of black holes and the bulges of their host galaxies \citep[e.g.][]{Magorrian_1998,Gebhardt_2000,  Haering_2004, Ferrarese_2000}, clear observational evidence for star formation quenching is still surprisingly scarce.  The actual link between these outflows and star formation rate quenching has been hard to establish observationally and has even led to some contrasting results. A few sources show evidence for suppression of star formation in regions of high velocity winds \citep{Cano-Diaz_2012,Brusa_2015b, Cresci_2015, Carniani_2016}. But other objects  show enhancement of star formation through AGN feedback which can be caused by compression of gas clouds by AGN outflows leading to that gravitational collapse \citep{Cresci_2015, Cresci_2015b}. Statistical analyses of the relationships between outflows and star formation have also been inconclusive. While some studies show that the most vigorous star formation is suppressed in the highest luminosity AGN \citep{Fabian_2012, Page_2012, Farrah_2012, Lanz_2015, Shimizu_2015}, other works involving more direct measurements of the wind velocity and energy, struggle to find the same evidence \citep{Veilleux_2013,Delvecchio_2015,Balmaverde_2016}. These attempts at finding conclusive observational proof that AGN may be able to quench star formation and regulate the host galaxies' growth have shown that this problem is highly complex where several different factors may play a role. Such factors may include AGN luminosity, AGN accretion rate, stellar mass of the galaxy, the spatial distribution of the star formation rate (central `bulge-like' star formation may be easier to curtail than extended `disk-like' star formation) or ISM properties. One of the key difficulties is the lack of reliable star formation diagnostics in the presence of a powerful AGN \citep{Zakamska_2016}.

In this paper, we focus on investigating the role of AGN feedback using one indicator for the outflow strength, the velocity width of the [OIII] emission line, and one indicator for the star formation activity of the host galaxy, its specific star formation rate. The forbidden [OIII] emission line traces the low-density gas ionized by the AGN and originates in the narrow-line regions surrounding the AGN. Being an optical emission line, it is easily observable from the ground for low to intermediate redshift AGN and traces the AGN-ionized gas even on galaxy-wide scales \citep{Liu_2013a}. We have therefore assembled a large sample of powerful AGN at intermediate redshift $0.2 < z < 1$ for which we were able to measure the relevant properties.

The paper is organized as follows. Section 2 describes the different AGN subsamples which our total sample is comprised of and the measurement of outflow strength, star formation rate and stellar mass of the host galaxy. In Section 3 we compare the different subsamples and analyze potential biases. The main results of our feedback tests are presented in Section 4 and we discuss them in the context of other works and galaxy evolution models in Section 5. Section 6 presents a summary and our conclusions. Throughout the paper we assume $H_0 = 71$ km s$^{-1}$ Mpc$^{-1}$, $\Omega_m = 0.27, \Omega_{\Lambda} = 0.73$. 

\section{Samples, observations and measurements}

\begin{figure*}
\begin{center}
\includegraphics[trim = 0cm 4cm 0cm 2.5cm, clip = true, scale = 0.6]{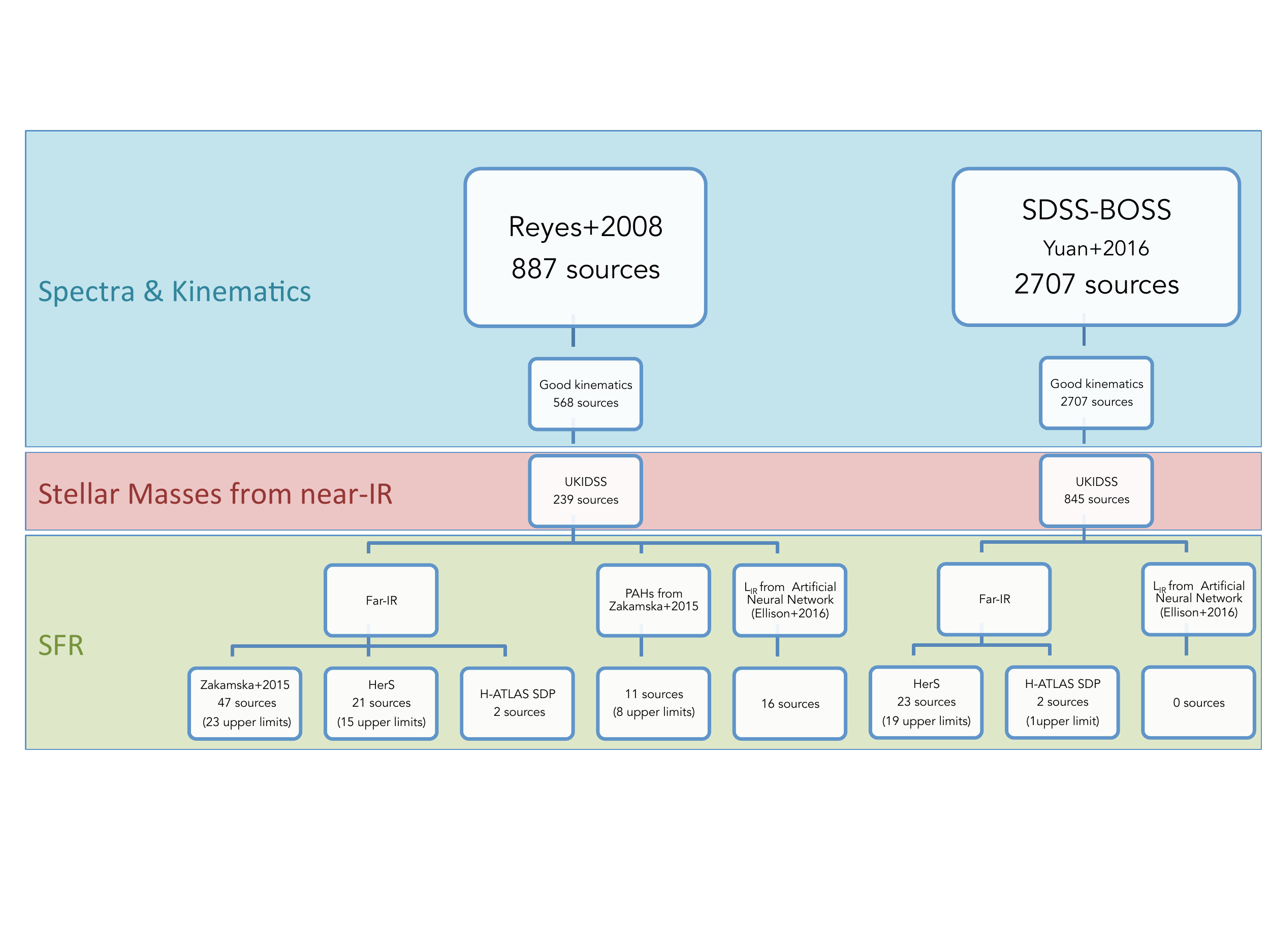}
\caption{Illustration of the selection process for the type-2 sources in our sample. We start out by using the two largest type-2 AGN catalogs at intermediate redshift, both originally selected from the SDSS and step-by-step narrow down the samples depending on which properties are measurable for the individual sources. We start off by measuring their ionized gas kinematics, which reduces the \citet{Reyes_2008} sample by about a third. We then cross-correlate all sources with near-IR catalogs from the UKIDSS survey. Detections in the near-IR are essential for deriving the stellar masses of the host galaxies. For the star formation rate measurements, we evaluate as a final step, which sources have detections in either the far-IR from \textit{Herschel} or \textit{Spitzer} (photometry from \citet{Zakamska_2016}, H-ATLAS or HerS survey), mid-IR spectroscopy for PAH measurements from \citet{Zakamska_2016} or star formation rate estimates from artificial neural network (ANN) analysis \citep{Ellison_2016}. This selection provides us with 122 sources (66 of which have upper limits on their star formation rates).}
\label{samples}
\end{center}
\end{figure*}

Luminous type-2 quasars in which the glow from the central black hole is obscured by dust allow us to observe their host galaxies which the quasar would otherwise outshine. They are therefore ideal targets for studying the quasars' effect on galaxy evolution. For this paper, we are making use of two of the largest intermediate-redshift ($z \sim 0.5$) type-2 AGN catalogs currently available: the type-2 AGN catalog published by \citet{Reyes_2008} and a new type-2 AGN catalog \citep{Yuan_2016} selected from the SDSS-III Baryon Oscillation Spectroscopic Survey \citep[BOSS,][]{Dawson_2013}. In the following sections, we describe how we select the final sample of type-2 AGN used in this paper and how we obtain or derive the relevant data to derive stellar masses, star formation rates and the properties of the ionized gas kinematics. In Figure \ref{samples} we illustrate the type-2 AGN selection process. We further supplement our sample with 10 type-1 AGN from \citet{Urrutia_2008, Urrutia_2012} (Section 2.3). 

\subsection{Type 2 AGN Sample I}

We primarily draw our type-2 AGN from the Sloan Digital Sky Survey \citep[SDSS, ][]{York_2000} candidates catalog presented by \citet{Reyes_2008}. The sample was selected based on emission line properties, such as emission line luminosities and ratios characteristic of ionization by a hidden AGN. For example, all sources are required to have a high [OIII]$\lambda$5007\AA/H$\beta$ ratio. This catalog contains 887 objects at $z < 0.8$.

\subsubsection{Gas kinematics}

Because the sample is selected from SDSS spectroscopy, optical spectra are available for all objects in the type-2 AGN catalog of \citet{Reyes_2008}. Using these spectra, we use non-parametric measurements that do not strongly depend on a specific fitting procedure to determine the width of the [OIII] emission line. We follow the measurement strategy presented in e.g. \citet{Zakamska_2014} and \citet{Liu_2013b}. Briefly, each profile is first fitted with multiple Gaussian components to determine the cumulative flux as a function of velocity:
\begin{equation}
\Phi(v) = \int_{-\infty}^{v} F_{v}(v') dv'
\end{equation}
For each spectrum, this definition is used to compute the line widths $W_{80}$ and $W_{90}$ that enclose 80\% and 90\% of the total flux, respectively. For a purely Gaussian profile, $W_{90}$ is closely related to the FWHM with $W_{80} = 1.4 \times$ FWHM, but the non-parametric velocity width measurements are more sensitive to the weak broad bases of non-Gaussian emission line profiles \citep{Liu_2013b}. We derive these quantities for 568 sources in the \citet{Reyes_2008} catalog with log([OIII]/L$_{\sun}) > 8.5$ where these measurements are most reliable. While we will also be quoting $W_{80}$ measurements for easier comparisons with other works, we will be using the $W_{90}$ measurements for quantitative analysis since it is very sensitive (even more so than $W_{80}$) to the broad bases of the emission lines.

\subsubsection{Stellar Masses} 

The emission of a galaxy in the rest-frame near-IR is less affected by recent star formation than emission in the optical. It can thus act as a good proxy for the well-established stellar population, i.e. the total stellar mass of the system \citep{Kauffmann_1998, Brinchmann_2000}. The UKIRT Infrared Deep Sky Survey \citep[UKIDSS,][]{Lawrence_2007} is a near-IR sky survey using WFCAM on the UK Infrared Telescope \citep[UKIRT,][]{Casali_2007} in Hawaii, that surveyed about 7500~deg$^2$ of the Northern Sky and is often considered to be the near-IR counterpart of the SDSS. We use the UKIDSS data release 10 and cross-correlate the UKIDSS catalogs with the 568 type-2 AGN from \citet{Reyes_2008} with reliable kinematic measurements using a pairing radius of 1~arcsec. This cross-correlation results in 239 matches with detections in at least one of the UKIDSS bands. We then utilize the SDSS optical photometric measurements of these galaxies ($ugriz$) and the near-IR photometry from UKIDSS to estimate the stellar masses $M_{\rm{stellar}}$ of these 239 type-2 AGN. We follow the spectral energy distribution (SED) fitting procedure described in \citet{Wylezalek_2016} using the Python package CIGALE \citep[Code Investigating GALaxy Emission,][]{Noll_2009}. Details on the fitting procedure and parameters can be found in \citet{Wylezalek_2016}. Briefly, we utilize the  \citet{Maraston_2005} stellar population models with a delayed star formation history $\rm{SFR(t)} = t \cdot \exp(-t/\tau)$. 

\begin{figure*}
\begin{center}
\includegraphics[trim =  0cm 0cm 0cm 1cm , clip = true, scale = 0.4]{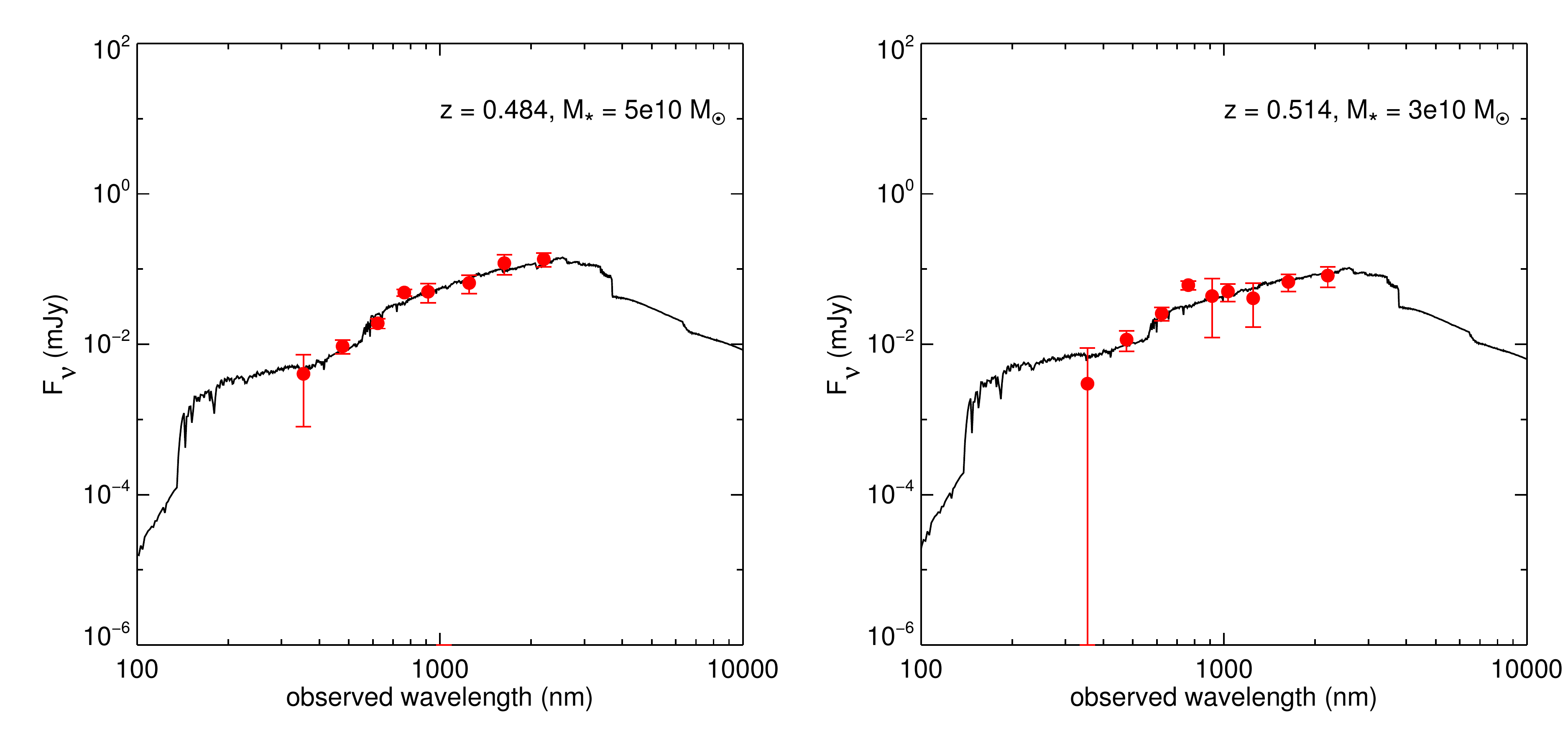}
\caption{Two example fits to the optical to near-IR photometry using CIGALE \citep{Noll_2009} to derive stellar masses of the host galaxies of our sources. \textit{Left:} Example spectral energy distribution and fit to a source from the \citet{Reyes_2008}+HerS subsample at $z = 0.484$. The $\chi^{2}_{\rm{reduced}} = 2.39$ and the derived stellar mass is $M_{\rm{stellar}} = 5\cdot10^{10}$~M$_{\odot}$. \textit{Right:} Example spectral energy distribution and fit to a sources from the BOSS+H-ATLAS subsample: $z = 0.514$, $\chi^{2}_{\rm{reduced}} = 4.6$, $M_{\rm{stellar}} = 3\cdot10^{10}$~M$_{\odot}$. }
\label{masses}
\end{center}
\end{figure*}

The optical to near-IR SEDs are well described by the models with a median reduced $\chi^{2} = 2.4$. Figure \ref{masses} shows the fitting results for two representative sources with derived stellar masses of  $M_{\rm{stellar}} = 5\cdot10^{10}$~M$_{\odot}$ and $M_{\rm{stellar}} = 3\cdot10^{10}$~M$_{\odot}$. We discuss the effect of the UKIDSS flux density limit on the stellar mass distribution in Section 3.3.

Since the sources in our sample are all AGN, at certain redshifts strong optical emission lines move into and out of broad-band optical filters which impacts the broad band fluxes. CIGALE does not account for these strong emission lines nor does it take into account the potentially contamination of AGN scattered light \citep{Obied_2016}. We therefore repeat the stellar mass derivation only using the $YJHK$ near-IR fluxes, which are not impacted by strong emission lines. Stellar masses derived from the near-IR data alone are on average identical with the derived masses from the combined optical and near-IR data with a mean deviation of about only 1\%. This confirms that the near-IR fluxes are more sensitive to the stellar mass than the optical fluxes since they best probe emission from the old stellar population. Optical broad band data in general has negligible impact on stellar mass estimates. 

We also note that in type-2 AGN the contribution of AGN emission components (e.g. contributions from a very hot dust component or direct emission from the central AGN) to the near-IR is negligible. The UKIDSS colors of our sources are consistent with colors of dominated by a stellar population whereas we would expect to observe redder colors were there a significant contribution of a hot dust component to the near-IR. Using a set of mock galaxies and the same SED fitting procedure as we do, CIGALE, \citet{Ciesla_2015} also show that the near-IR SEDs of type-2 AGN are not contaminated by non-stellar emission and that stellar masses can be recovered with high confidence. This is in agreement with results presented in \citet{Zakamska_2004} who demonstrate that the near-IR $J-K_{s}$ colors of type-2 AGN are consistent with being dominated by the light from the host galaxies. Although we only focus on type-2 AGN in this paper, we note that the situation can be vastly different in intermediate-type or type-1 AGN \citep{Ciesla_2015} or at mid-IR wavelengths $\lambda > 3\mu$m where even $\sim 35$\% of type-2 AGN show contributions from dust emission \citep{Yuan_2016}.

We conclude that strong optical emission lines impacting broad band fluxes in the optical or contribution from direct AGN or torus emission to the near-IR do not have any impact on the accuracy of our stellar mass estimates. We will therefore be using near-IR only derived stellar masses for the subsequent analysis in this paper.  

\subsubsection{Star Formation Rates} 

Measuring star formation rates for AGN hosts is a challenging task. The most common techniques used for star-forming galaxies include optical and mid-infrared emission lines, polycyclic aromatic hydrocarbon (PAH) features and mid- and far-infrared luminosities \citep{Kennicutt_1998a}. In AGN, however, all these diagnostics can be severely affected by ionizing radiation by the AGN which introduces a large scatter to these calibrations. We utilize the least biased techniques to measure SFRs for our AGN in this paper. 

\vspace{0.3cm}
\textbf{Far-IR:} Far-IR emission is produced in galaxies when UV/optical photons from young, massive stars heat up interstellar dust particles which then re-emit at longer wavelengths. The IR luminosity of a galaxy is thus directly related to its instantaneous obscured star formation rate. While technically the obscured and the unobscured star formation rate, which can be measured directly from the emission in the UV, have to be added to obtain the total star formation rate, various analyses have shown that the unobscured rates contribute a negligible fraction to the total star formation rate in luminous type-2 AGN \citep{Lacy_2007, Obied_2016}. The obscured star formation rates derived from the far-IR emission are therefore a good measure for the total star formation rate of luminous, intermediate redshift type-2 quasar hosts.

In AGN, heating of dust grains by the AGN can have significant impact on the total IR luminosity, usually measured by integrating the IR SED from rest-frame $8\mu$m to 1000$\mu$m. AGN-heated dust is much more compact than the star-heated dust which is distributed on scales of the host galaxy, so that dust temperatures associated with AGN heating are significantly higher \citep{deGrijp_1985, Kirkpatrick_2014}. Emission from AGN-heated dust therefore mostly affects the IR SED at $\lambda_{\rm{rest}} < 100 \mu$m. \citet{Wylezalek_2016} show that star formation rates derived by decomposing the far-IR spectral energy distribution (SED) into contributions from AGN and SF are consistent within a factor of two with star formation rates derived through single band far-IR ($\lambda_{\rm{rest}} > 100 \mu$m) measurements, where contribution from the AGN is negligible \citep{Zakamska_2016}. There are some exceptions where the SED is dominated by the AGN even at far-IR  ($\lambda_{\rm{rest}} > 100 \mu$m) wavelengths, especially when the AGN is luminous \citep{Hony_2011, Sun_2014} and SFRs are low. But in these rare cases, the measured far-IR luminosity would imply SFRs $\sim 10$ M$_{\odot}$/yr if all the far-IR emission were assumed to be due to SF despite being dominated by the AGN. Given the typical sensitivity of our \textit{Herschel} data, we would not be sensitive to the far-IR emission of such rare AGN. In such cases, we would report an upper limit corresponding to the flux limit of the far-IR observations which is typically higher than the SFR in these rare AGN. Therefore, even in the cases where far-IR is not due to star formation, due to our measurement strategy we would report a conservative upper limit.

For this work, we have conducted an extensive search for far-IR measurements from the \textit{Herschel Space Observatory} for the 239 sources in \citet{Reyes_2008} with UKIDSS data. We are using the following data sets:
\begin{itemize}
\item \citet{Zakamska_2016} recently compiled available far-IR photometry at 160$\mu$m taken with the Multi-band Imaging Photometer for Spitzer \citep[MIPS,][]{Rieke_2004} onboard the \textit{Spitzer Space Telescope} for the type-2 AGN presented in \citet{Reyes_2008}. This sample is heterogenous, consisting of sources that were observed for different reasons (e.g. high $L_{\rm{[OIII]}}$, radio sources) or serendipitously. Furthermore, far-IR photometry from the \textit{Herschel Space Observatory} is available for subsamples of the \citet{Reyes_2008} AGN catalog from two programs (PIs: Zakamska, Ho), both selected on $L_{\rm{[OIII]}}$ criteria. Using a variety of templates and scaling relations, \citet{Zakamska_2016} derive far-IR luminosities for these sources with an accuracy of about 0.2~dex. In case of weak far-IR or non-detections, upper limits are reported. We follow \citet{Kennicutt_1998} to convert $L_{\rm{IR}}$ to SFRs. For 107 of those \textit{Spitzer} and \textit{Herschel} sources, reliable kinematic measurements are available from SDSS spectra, out of which 47 sources (including 23 upper limits on SFR) are detected in the near-IR with UKIDSS. We include these 47 sources in our sample presented in this paper.

\item The \textit{Herschel} Stripe 82 survey \citep[HerS,][]{Viero_2014} utilized the SPIRE instrument aboard the \textit{Herschel Space Observatory} to obtain 79 deg$^2$ of contiguous deep imaging at 250, 350 and 500 $\mu$m. HerS was designed to overlap with a number of existing and planned galaxy surveys in the SDSS `Stripe 82' field. The HerS catalog has been made public and contains 32,815 band-matched sources with significance greater than 5$\sigma$ ($S_{250\mu m} >$ 31~mJy). We have cross-correlated all 239 type-2 AGN from \citet{Reyes_2008} with UKIDSS detections with this catalog using a pairing radius of 6 arcsec (roughly half the size of the SPIRE beam at $250\mu$m), resulting in 6 matches which are all within 3 arcsec of the original SDSS position. Additionally, we include sources which fall into the field of HerS but are not detected in the HerS catalog. This adds another 15 sources to the sample with upper limits on IR luminosity. 

To estimate SFRs from these far-IR data, we first estimate the total IR luminosity utilizing the method presented in \citet{Symeonidis_2008}. Using a sample of over 40 infrared-luminous galaxies at $0.1 < z < 1.2$, \citet{Symeonidis_2008} show that the total infrared luminosity can be estimated using single band detections at rest-frame 24, 70 or 160 $\mu$m with a fairly small scatter of 0.18, 0.11 and 0.23 dex rms, respectively. 

For our sources, we assume that the far-IR emission measured by \textit{Herschel} is dominated by the contribution from star-formation \citep{Zakamska_2016} and that the shape follows that of a modified black body with a spectral emissivity index $\beta = 1.5$ and temperature $T=45$K. Using those assumptions we derive $K$-corrections for the $250\mu$m detections from HerS to compute rest-frame 160$\mu$m luminosities. We then use the relations presented in \citet{Symeonidis_2008} to derive total infrared luminosities $L_{\rm{IR}}$ and then follow \citet{Kennicutt_1998} to convert $L_{\rm{IR}}$ to SFRs. Given the uncertainties in the SPIRE data and the monochromatic far-IR luminosities to total IR luminosities conversion, we estimate the uncertainties on the SFRs to be of order 0.4~dex. For the HerS non-detections, we use the detection limits of the catalog, 31 mJy, to derive upper limits on their SFRs.

\item 
H-ATLAS is a large extragalactic sky survey which used the PACS and SPIRE instruments onboard the \textit{Herschel Space Observatory} to observe over 500 deg$^{2}$ split between six fields. Data and and band-matched catalogs are publicly available only for the $4 \times 4 $~deg$^2$ science demonstration phase (SDP) field. We cross-match the \citet{Reyes_2008} type-2 AGN catalog with this public H-ATLAS catalog using a pairing radius of 6 arcsec and the UKIDSS near-IR catalog, resulting in 2 matches which are both within 3 arcsec of the original SDSS position. Although both sources are undetected at $160\ \mu$m, we use their detections at 250$\mu$m to infer SFRs following the the same method as described for the HerS sources. No additional sources from \citet{Reyes_2008} lie in the H-ATLAS SDP field.
\end{itemize}

\vspace{0.3cm} 
\textbf{PAH features:} The mid-IR spectra of many galaxies are dominated by emission features at 3.3, 6.2, 7.7, 8.6 and 11.2 $\mu$m attributed to Polycyclic Aromatic Hydrocarbons (PAHs) and are typically used as a star formation rate indicator \citep{Peeters_2004, Calzetti_2011}. \citet{Zakamska_2016} searched the $\textit{Spitzer}$ Heritage Archive for mid-IR spectra for sources in the \citet{Reyes_2008} catalog and found matches for 46 sources. Those sources were targeted by different groups with the \textit{Spitzer Space Telescope} Infrared Spectrograph \citep[IRS;][]{Houck_2004} as type-2 quasar candidates selected on [OIII] luminosity and/or on infrared, optical or X-ray properties. In general, those 46 sources represent well the \citet{Reyes_2008} sample of type-2 quasars \citep{Zakamska_2016}. Carefully investigating and testing various PAH-to-SFR calibrations, \citet{Zakamska_2016} finally derive SFRs (or upper limit on SFRs) for those sources. 11 of the type-2 AGN with PAH derived SFRs (of which eight are upper limits) are detected by UKIDSS and we include those 11 sources in our work. 

\vspace{0.3cm}
\textbf{$L_{\rm{IR}}$ from Artificial Neural Networks:} \citet{Ellison_2016} recently used a set of 1136 galaxies in HerS with known $L_{\rm{IR}}$ to train an artificial neural network (ANN). This ANN requires a set of optical measurements (a mix of line luminosities, optical magnitudes and colors) to reliably predict  $L_{\rm{IR}}$ with a scatter of 0.23 dex and no systematic offset. Most importantly, \citet{Ellison_2016} showed that their ANN performs equally well both on star-forming galaxies and AGN. Using the ANN, they predict $L_{\rm{IR}}$ for over 300,000 galaxies in the SDSS and make their catalog publicly available. We cross-correlate this ANN catalog with the \citet{Reyes_2008} type-2 AGN catalog and UKIDSS and find 30 matched sources. We further restrict ourselves to sources with  the associated `error' $\sigma_{\rm{ANN}} < 0.3$, the scatter between the network outputs. This results in a final ANN sample size of 16 sources. We then convert the $L_{\rm{IR}}$ reported by \citet{Ellison_2016} to SFRs following \citet{Kennicutt_1998}. 

\subsection{Type 2 AGN Sample II}

The second type-2 AGN sample we utilize in this paper is a new type-2 AGN catalog selected from the SDSS-BOSS survey \citep{Eisenstein_2011, Dawson_2013}. Selection was performed in a similar fashion to \citet{Reyes_2008}, based on a combination of various line luminosity cuts and emission line ratios. The selection and source catalog are presented in \citet{Yuan_2016}. This BOSS sample contains 2707 objects with a median redshift and standard deviation of $z = 0.47 \pm 0.13$. We find 845 matches with the UKIDSS near-IR catalog which allows us to derive stellar masses for those 845 sources. Using the same far-IR surveys as above, we cross-correlate the BOSS-UKIDSS sources with the HerS, H-ATLAS and the ANN catalogs and find 4,1 and zero matches, respectively. Additionally, we include BOSS-UKIDSS sources which fall into the field of HerS and H-ATLAS SPD, but are detected at less than 3$\sigma$ and 5$\sigma$ significance, the quality cuts chosen by the HerS and H-ATLAS teams, respectively. This adds another 19 and 1 sources to the sample. We use the detection limits of the catalogs (31 mJy and 7 mJy, respectively) to derive upper limits on their SFRs. 

\subsection{Type 1 AGN sample}

In addition to the two larger type-2 AGN samples from \citet{Reyes_2008} and BOSS, we further include 10 out of 13 sources from the red quasar sample presented in \citet{Urrutia_2008} \citep[see also][]{Glikman_2012, Urrutia_2012}. Using red J-K colors, \citet{Urrutia_2008} found a large sample of red quasars that were followed up spectroscopically at Keck and IRTF in subsequent years \citep{Glikman_2012}. These quasars are thought to be red in the optical/near-IR due to moderate dust reddening ($A_{V} \sim 1- 5$ mag),  but are highly luminous which is clear from their IR emission. A representative sample of 13 sources with $0.4 < z < 1$ was observed with the Advanced Camera for Surveys (ACS) on the \textit{Hubble Space Telescope} (HST) and also with MIPS and IRS onboard the \textit{Spitzer Space Telescope}. We have analyzed 10 out of those 13 red quasars using the spectra provided by \citet{Glikman_2012} \citep[kinematic decomposition presented in][]{Brusa_2015a} with sufficient S/N and derived the narrow-line region kinematic parameters described in Section 2.1.1. 

The challenge with deriving host galaxy parameters for this sample of objects, which are type-1 AGN, is that despite significant reddening the quasar outshines the galaxy in the optical and strongly contaminated the near-IR, making simple SED fitting for deriving stellar masses impossible. Using the high-resolution ACS images taken with the HST, \citet{Urrutia_2008} carefully perform quasar point-source extraction and host galaxy fitting and derive host galaxy parameters such as morphology, optical magnitudes and nuclear-to-host-galaxy ratios. \citet{Urrutia_2008} show that the $g'-I_{\rm{C}}$ colors of the host galaxies of these type-1 AGN are not particularly blue and therefore show little unobscured star formation. They also calculate absolute B-band magnitudes $M_{B}$ of the hosts adopting a 1 Gyr post starburst model from \citet{Bruzual_2003} and compare them with the luminosity break $L^{*}$of the galaxy luminosity function at the median redshift of the sample, $z \sim 0.7$, from \citet{Ilbert_2006}. Most of the host galaxies are found to be around or above $L^{*}$ which corresponds to a stellar mass of $M^* \sim 10^{10.9} M_{\odot}$ \citep{Ilbert_2010}. To calculate the mass of the host galaxies in the \citet{Urrutia_2008} sample, we use their host luminosities and convert them to masses using the $M^*/L^*$ ratio. 

\citet{Urrutia_2012} present far-IR observations with MIPS at $24,\ 70$ and $160\mu$m for the same sample of red quasars. Combining their multi-wavelength photometry ranging from the UV to the infra-red they perform SED fitting and decompose the far-IR emission into contributions by AGN-heated dust and star-formation-heated dust and report the far-IR luminosities associated with these two heating mechanisms. We follow again \citet{Kennicutt_1998} to convert their $L_{\rm{IR}}$ to SFRs. 
\newline\newline
Combining all these subsamples, we have now accumulated a sample of 132 type-2 and type-1 AGN, for which we derived narrow-line region kinematic properties, stellar masses and star formation rates. Since all sources from the type-2 subsamples \citep{Reyes_2008, Yuan_2016} have their masses derived from SDSS+UKIDSS photometry and narrow-line kinematics derived from their SDSS spectra, we will refer to the various subsamples based on how their SFR was derived: NZ16, S82 (HerS), H-ATLAS, PAH, ANN, BOSS S82 (HerS), BOSS H-ATLAS. We will refer to the type-1 AGN subsample as the Urrutia objects. 

\section{Comparison of subsamples}
 
One of the challenges with working with such a heterogenous dataset is to make sure to not introduce any biases in the analysis. For that reason, we investigate here how well the different subsamples used in this paper cover the range in parameters that are essential for this work. 

We point out that in this work we focus on high-luminosity AGN, most of them above the quasar threshold ($L_{\rm bol}=10^{45}$ erg/sec). The reasons for this choice are two-fold. First, in such sample we can safely assume that the mid-IR emission is dominated by the AGN, not star-formation, making it possible to have both AGN and SFR measurements from scant SED coverage. This assumption is confirmed by a lack of correlation between mid-IR and far-IR measurements. In lower-luminosity objects a full SED decomposition would be required to obtain the AGN luminosity. Second, theoretical models \citep{Zubovas_2012b} suggest that the AGN needs to provide sufficient luminosity to be able to push the gas out of the galactic potential. This `threshold' nature of AGN feedback was recently also suggested by molecular gas \citep{Veilleux_2013} and radio observations \citep{Zakamska_2014}.

\subsection{Redshift coverage}

\begin{figure}
\begin{center}
\includegraphics[trim =  3cm 0.5cm 0cm 1cm , clip = true, scale = 0.35]{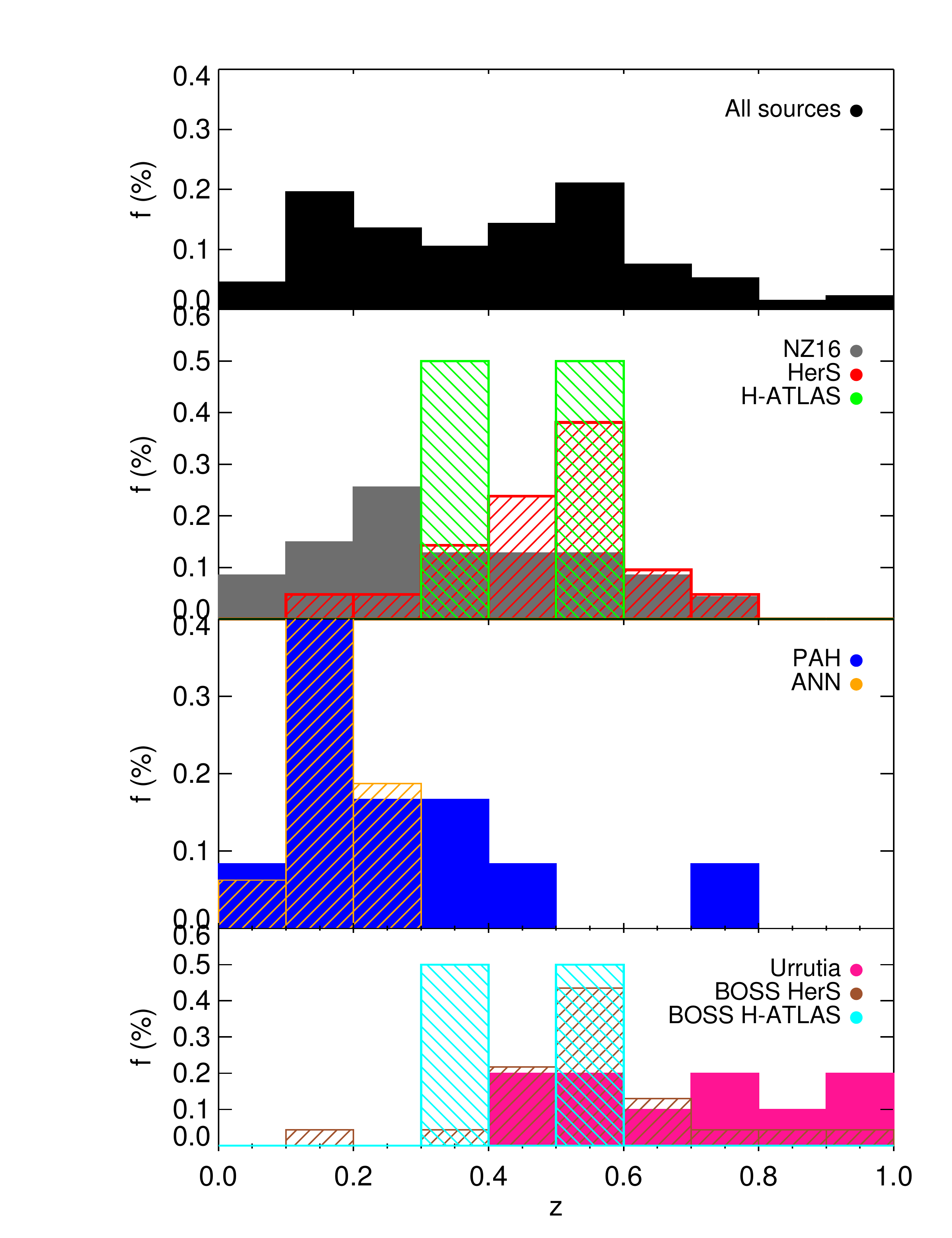}
\caption{Redshift distribution of all sources used in this paper and of the different subsamples, split up in three panels for clarity. The Urrutia subsample probes the highest redshift objects, while the ANN subsample probes sources at the lower end of the overall redshift distribution. Overall the redshift range probed is pretty broad with a mean of $z = 0.4 \pm 0.2$.}
\label{z_histo}
\end{center}
\end{figure}

\begin{figure*}
\begin{center}
\includegraphics[trim =  3cm 0cm 0cm 1cm , clip = true, scale = 0.35]{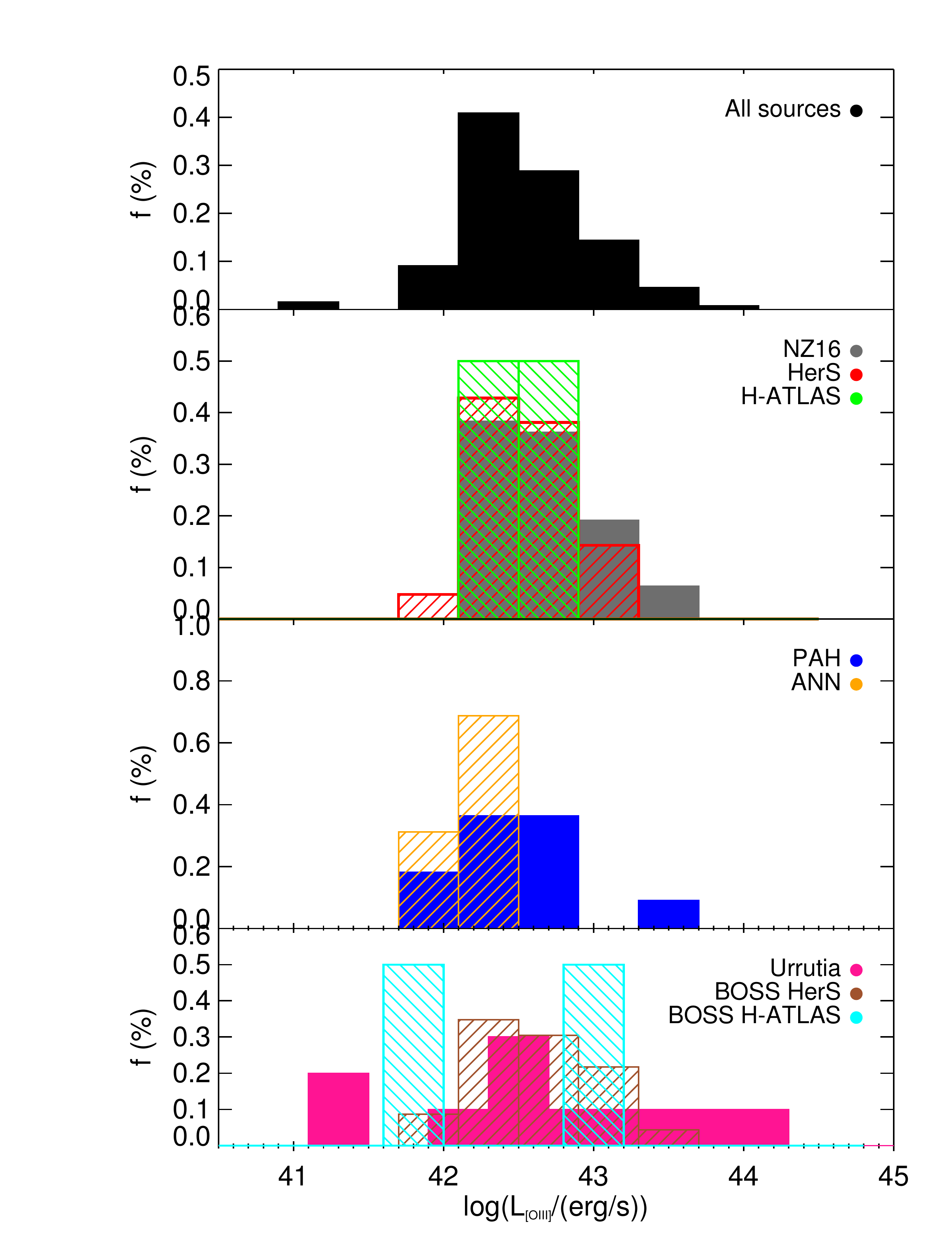}
\includegraphics[trim =  3cm 0cm 0cm 1cm , clip = true, scale = 0.35]{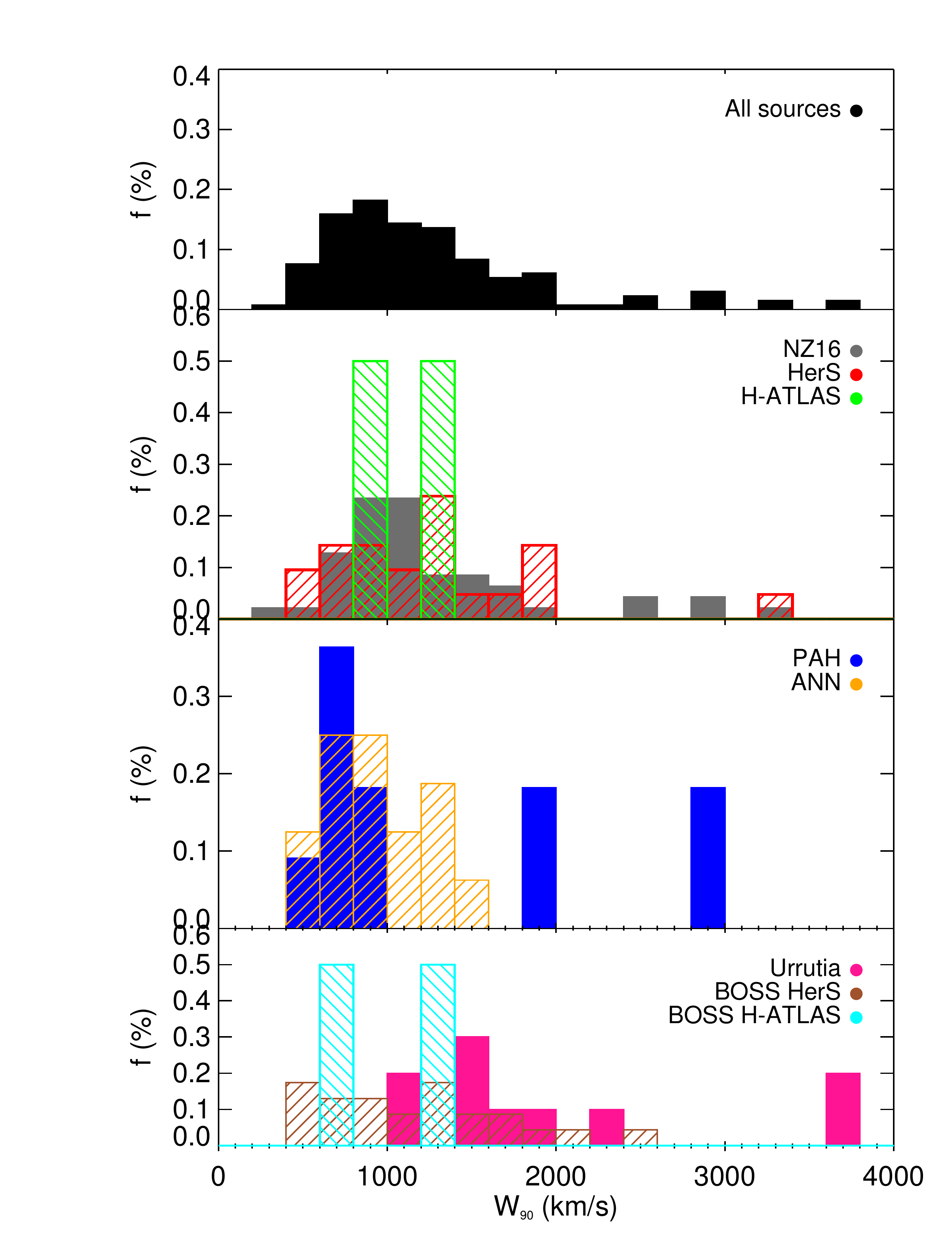}
\caption{$L_{\rm{[OIII]}}$ (left) and $W_{90}$ (right) distribution of all sources used in this paper and of the different subsamples, split up in three panels for clarity, respectively. The NZ15, HerS, H-ATLAS and PAH subsamples were, among other criteria, originally selected based on their $L_{\rm{[OIII]}}$ and therefore make up most of the high luminosity end of the overall luminosity distribution. The other subsamples nicely complement them at the lower luminosity end. The range in luminosities spans $L_{\rm{[OIII]}} = 10^{41.5}-10^{44}$~erg~s$^{-1}$. The median $W_{90}$ is $\sim1100$~km~s$^{-1}$, which is too high to be explained by gas rotation in a galactic disk. High values of $W_{90}$ are indicative of gas out of the dynamical equilibrium with the galaxy, likely in an outflow driven by the AGN. All subsamples probe a wide range of $W_{90}$ values.}
\label{oiii_histo}
\end{center}
\end{figure*}

While the peak of quasar activity and galaxy growth, i.e. star formation rate, where mergers and galaxy interactions were more common and major mergers may have played a significant role in the triggering of quasars happened at around $z \sim 2$, one of the best and most commonly used indicators for AGN-excited gas on large scales are the forbidden lines [OIII]$\lambda\lambda$4959,5007 which trace low-density gas. At $z > 1$, observing these lines becomes increasingly difficult using ground-based observatories. We are therefore focusing on AGN with $0.1 < z < 1$ where large spectroscopic data sets including the observations in this paper are available. The mean redshift of all our subsamples combined and its standard deviation is $z = 0.4 \pm 0.2$, where the Urrutia sample probes the highest redshift population ($z = 0.7 \pm 0.2$) and the ANN subsample the lowest redshift population ($z = 0.20 \pm 0.05$). Figure \ref{z_histo} shows the redshift distribution for the different subsamples, where we have split up the distributions in three panels for clarity. The overall redshift distribution in relatively flat between $z = 0.1$ and $z = 0.6$ with only a few sources at the high and low redshift end. 

\subsection{[OIII] luminosity and line width ($W_{90}$) coverage}

Although $L_{\rm{[OIII]}}$ can be affected by extinction, most likely from dust in the narrow-line region, [OIII] luminosities are a good indicator of total bolometric AGN luminosity \citep{Reyes_2008, LaMassa_2010}. The [OIII] luminosities of the sources in this paper cover the range of $L_{\rm{[OIII]}} = 10^{41.5}-10^{44}$~erg~s$^{-1}$ (Figure \ref{oiii_histo}) which is similar to the typical range of $L_{\rm{[OIII]}}$ as measured from the distribution of $L_{\rm{[OIII]}}$ luminosities in the \citet{Reyes_2008} sample in the same approximate redshift range. This shows that the sources in our work well represent the overall population of optically selected AGN at $0.1 < z < 1$. These line luminosities correspond roughly to a total bolometric luminosity of $L_{\rm{bol}} = 10^{44.5-47}$~erg~s$^{-1}$ \citep{Reyes_2008}. 

Because the sources in the NZ15, HerS, H-ATLAS and PAH subsamples were partly selected on their $L_{\rm{[OIII]}}$ luminosities, it is of no surprise that these subsamples do not contribute significantly to the low luminosity end of the overall luminosity distribution. The sources in the ANN and Urrutia samples were selected on their robustness of mass measurements (ANN) and their near-IR properties (Urrutia) and therefore tend to lower [OIII] luminosities, nicely supplementing the higher [OIII] luminosity subsamples. 

The [OIII] velocity width $W_{80}$ or $W_{90}$ measures the velocity dispersion of the ionized gas and are a good proxy for the radial outflow velocity of the gas. \citet{Liu_2013b}, for example, shows that $W_{80} \sim 1.5 v_{\rm{outflow}}$. The median $W_{80}$/$W_{90}$ of the full sample is $\sim 770/1100$~km~s$^{-1}$, ranging up to velocity widths as high as $2900/3600$~km~s$^{-1}$ (Figure \ref{oiii_histo}). Spatially integrated line-of-sight velocity dispersion of $250$~km~s$^{-1}$ (corresponding to $W_{80} \sim 640$~km~s$^{-1}$ or $W_{90} \sim 800$~km~s$^{-1}$) are typically not exceeded by the ionized gas in stellar disks. Even in morphologically disturbed high-redshift ultraluminous galaxies, models \citep{Lehnert_2009} and observations \citep{Harrison_2012, AlaghbandZadeh_2012} show that integrated line-widths are not expected to exceed $W_{80}$/$W_{90} \sim 500/650$~km~s$^{-1}$. The wind velocities in the sources in our sample are significantly higher and show that the gas cannot be in dynamical equilibrium with the host galaxy potential. Such observed velocity dispersions typically exceed the escape velocity of the galaxy and are clear indicators of a wind where gas is being removed from the host galaxy. Starburst-driven outflows typically do not exceed $W_{90}$ values above 1000~km/s \citep{Hill_2014} \citep[see also][]{Rupke_2013, Heckman_2015}, the outflows we observe here are therefore very likely driven by the central AGN. 

One of the goals of this paper is to find out if and how strongly the galactic wind couples to the gas from which stars form and to which extent this coupling might influence the star formation in the galaxy. All data sets used in this work cover a wide range of velocity dispersions so that any possible biases are expected to be minimal.

\subsection{Host stellar mass coverage}

A critical measurement for the work presented here is achieving good estimations of the stellar mass for our sources. As described in Section 2.1.2, we deploy the same mass estimation algorithm to all sources except the Urrutia sources. In the case of the Urrutia subsample, HST imaging allowed these authors to separate the host emission from that of the quasar. But in general, in type-1 AGN which can dominate optical and NIR emission measuring the stellar mass is very difficult. This presents a major obstacle for repeating our investigation with its focus on the specific star formation rates for a larger sample of type-1 AGN.

\begin{figure}
\begin{center}
\includegraphics[trim =  3cm 0cm 0cm 1cm , clip = true, scale = 0.35]{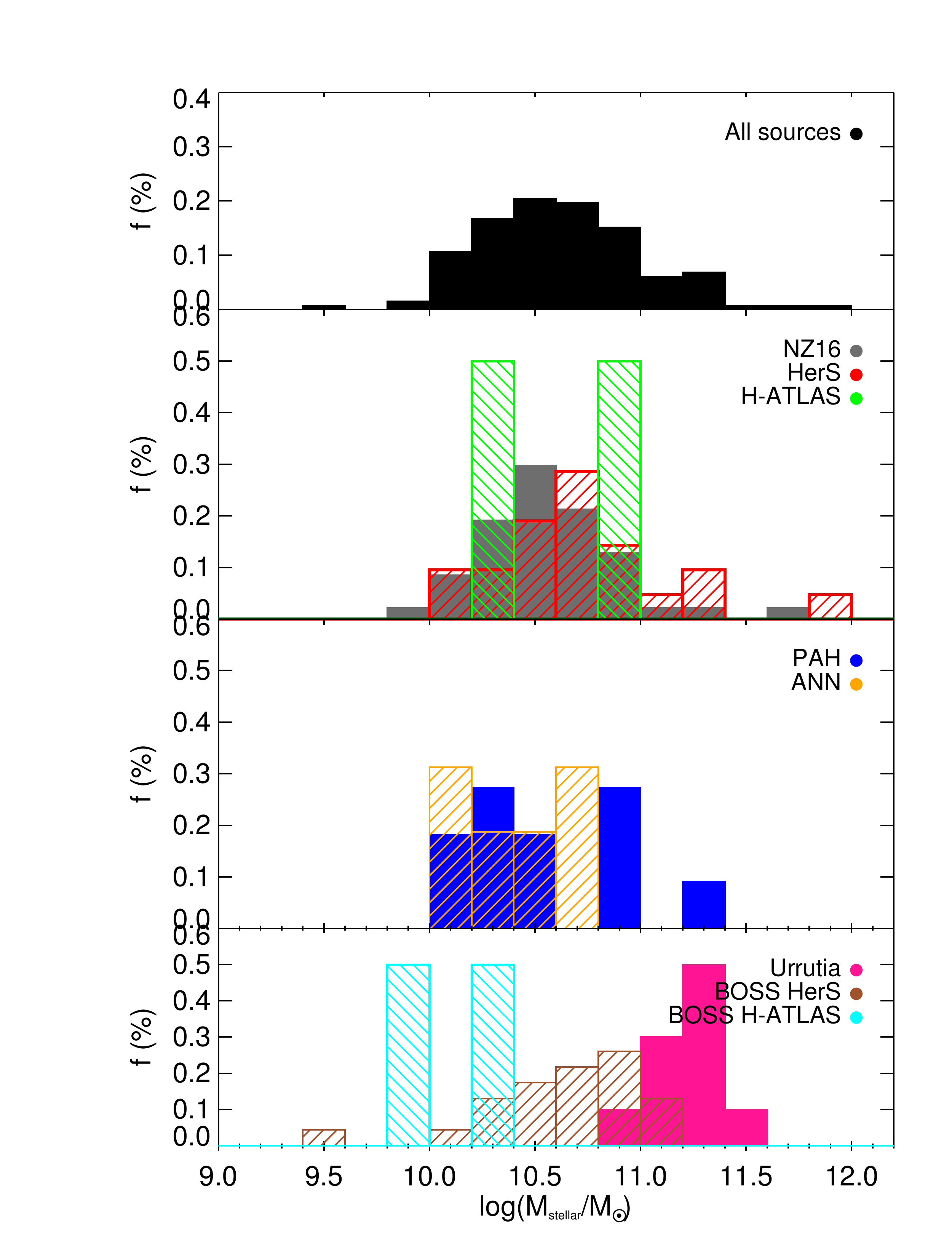}
\caption{Stellar mass distribution of all sources used in this paper and of the different subsamples, split up in three panels for clarity. The Urrutia subsample probes the objects at the highest mass with log($M/M_{\odot}) = 11.2 \pm 0.2$, which is likely a result of the initial selection process. The other subsamples probe a slightly lower and broader mass range, log$(M/M_{\odot}) = 10.5 \pm 0.4$, related to the more diverse and optical emission-line based target selection.}
\label{mass_histo}
\end{center}
\end{figure}

The mean mass for our total sample is log$(M_{\rm{stellar}}/M_{\odot}) = 10.5 \pm 0.4$, which is about 0.3 dex below $M^{*}$ for star-forming galaxies in the same redshift range \citep{Muzzin_2013}. The mean mass of the Urrutia subsample is significantly higher than that of the overall population (log($M_{\rm{stellar}}/M_{\odot}) = 11.2 \pm 0.2$, Figure \ref{mass_histo}) which is most likely a result of the initial selection of that sample. Part of the selection included a detection in the FIRST radio survey \citep{Becker_1995} and their radio fluxes are such that they can be classified as radio-intermediate/radio-loud sources which tend to reside in more massive host galaxies \citep{Best_2005, Seymour_2007}. Additionally, the Urrutia subsample is the on average highest-redshift subsample enhancing this selection bias.   


The survey depth of the UKIDSS survey, K = 18.3 mag, determines the lower mass limits for the galaxies studied here. At $z = 0.2$, the lower limit on $M_{\rm{stellar}} \sim 5\times10^9$~M$_{\sun}$, whereas at $z = 0.5$, the lower limit on $M_{\rm{stellar}} \sim 2\times10^{10}$~M$_{\sun}$ \citep{Kauffmann_1998}. This limit is reflected in the minimum stellar masses of the different subsamples in this work. We discuss the effect of this redshift-dependent mass limit on our results in Section 4.

\subsection{Star formation rate coverage}

Star formation rate in our galaxies is primarily measured using far-IR luminosity measurements from \textit{Herschel} and \textit{Spitzer}. The main limitation for actual star formation rate measurements are therefore the depths of the surveys used, which are 7~mJy for the H-ATLAS survey and 30~mJy for the HerS survey. For a galaxy at $z = 0.5$ these limits translate to detection limits on star formation rates of $20$ and $70$~M$_{\odot}$/year, respectively. We further include upper limits on star formation rate for all sources which fall into the covered area of the relevant far-IR surveys but are formally not detected. For the distribution of SFR measurements 

\begin{figure}
\begin{center}
\includegraphics[trim =  3cm 0cm 0cm 1cm , clip = true, scale = 0.35]{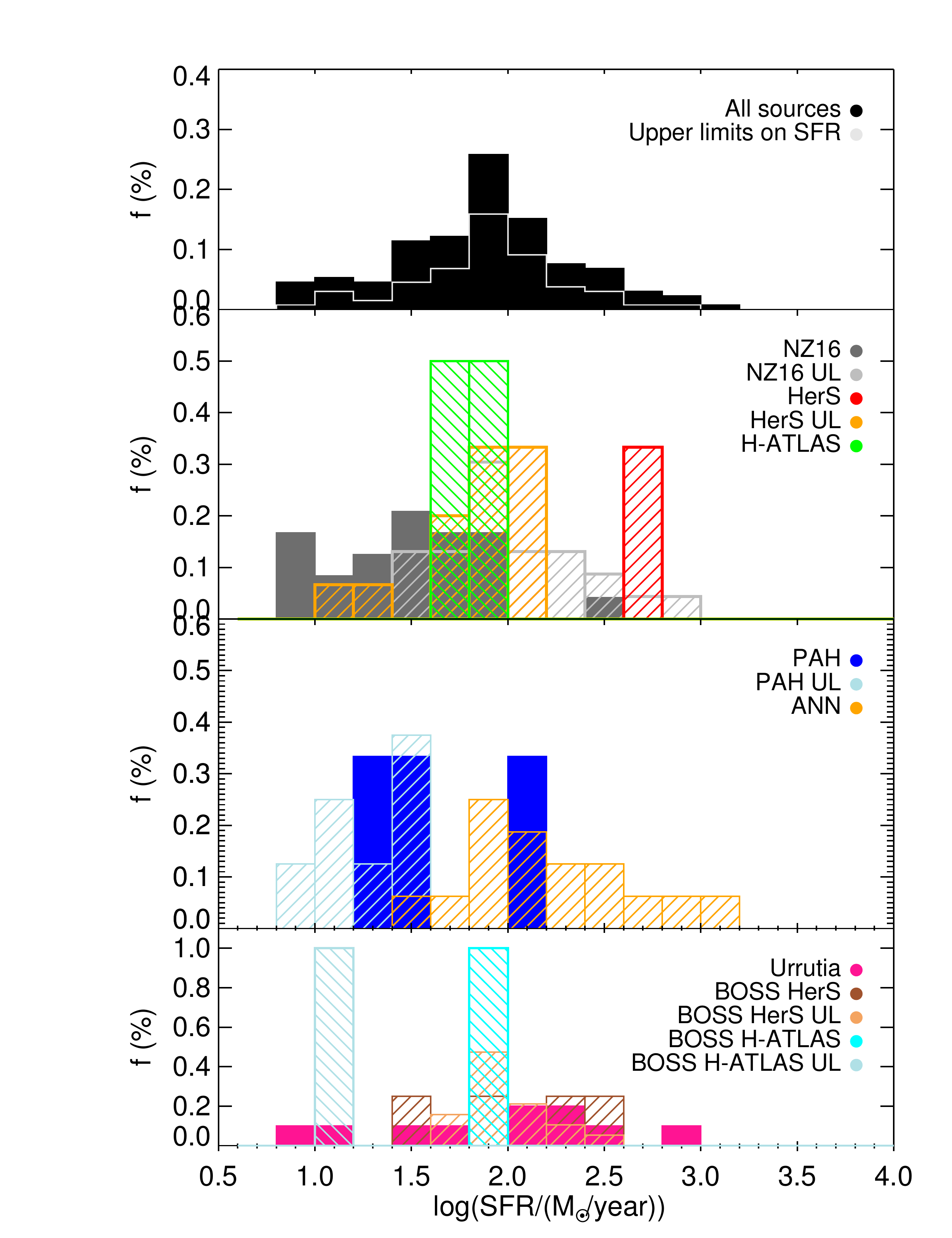}
\caption{Star formation rate distribution of all sources used in this paper and of the different subsamples, split up in three panels for clarity. The abbreviation `UL' is used when only upper limits on SFR are available.}
\label{sfr_histo}
\end{center}
\end{figure}

\begin{figure}
\begin{center}
\includegraphics[trim =  5.2cm 1cm 45cm 39.5cm , clip = true, scale = 0.195]{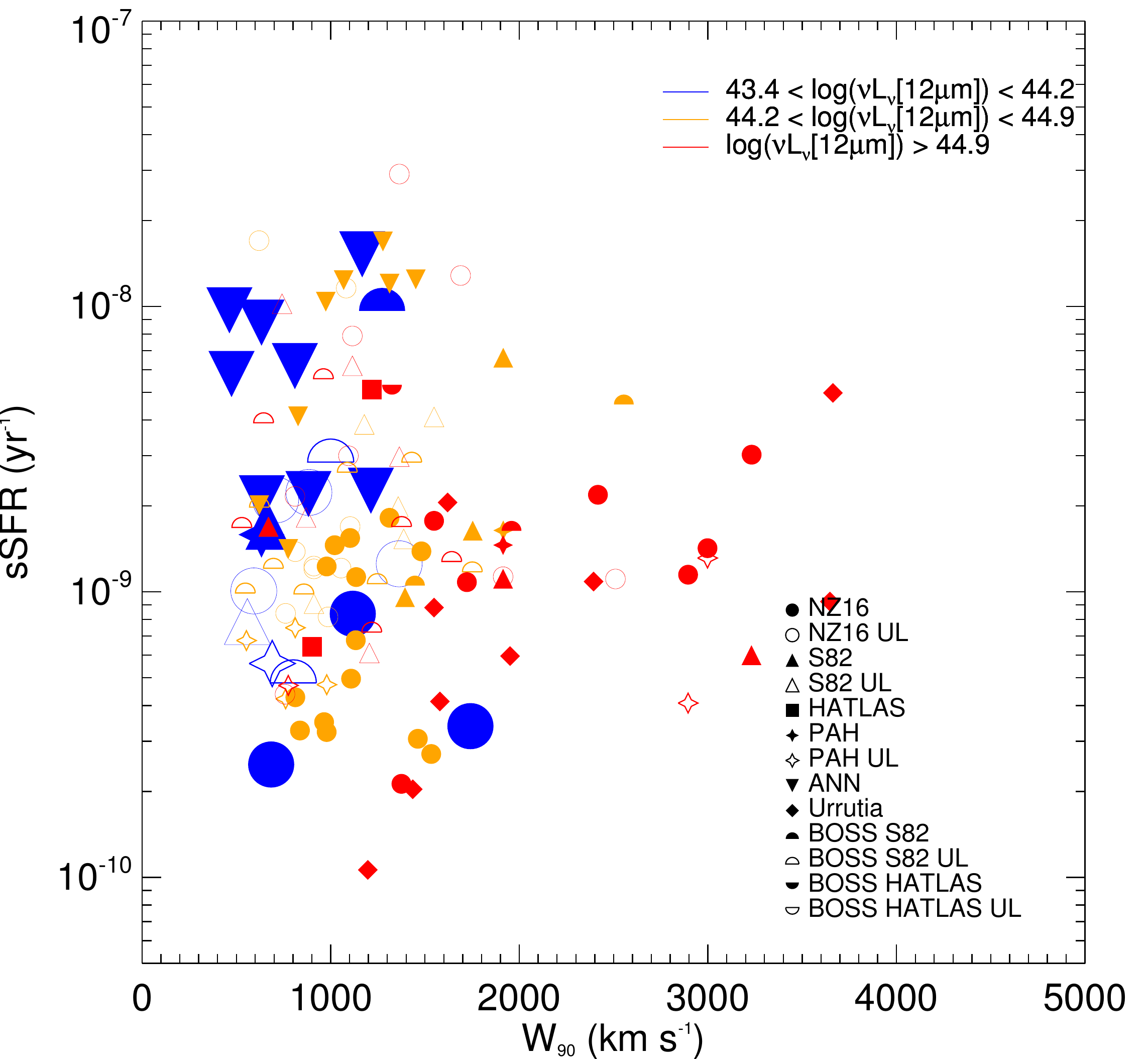}
\includegraphics[trim =  5.2cm 1cm 45cm 39.5cm , clip = true, scale = 0.195]{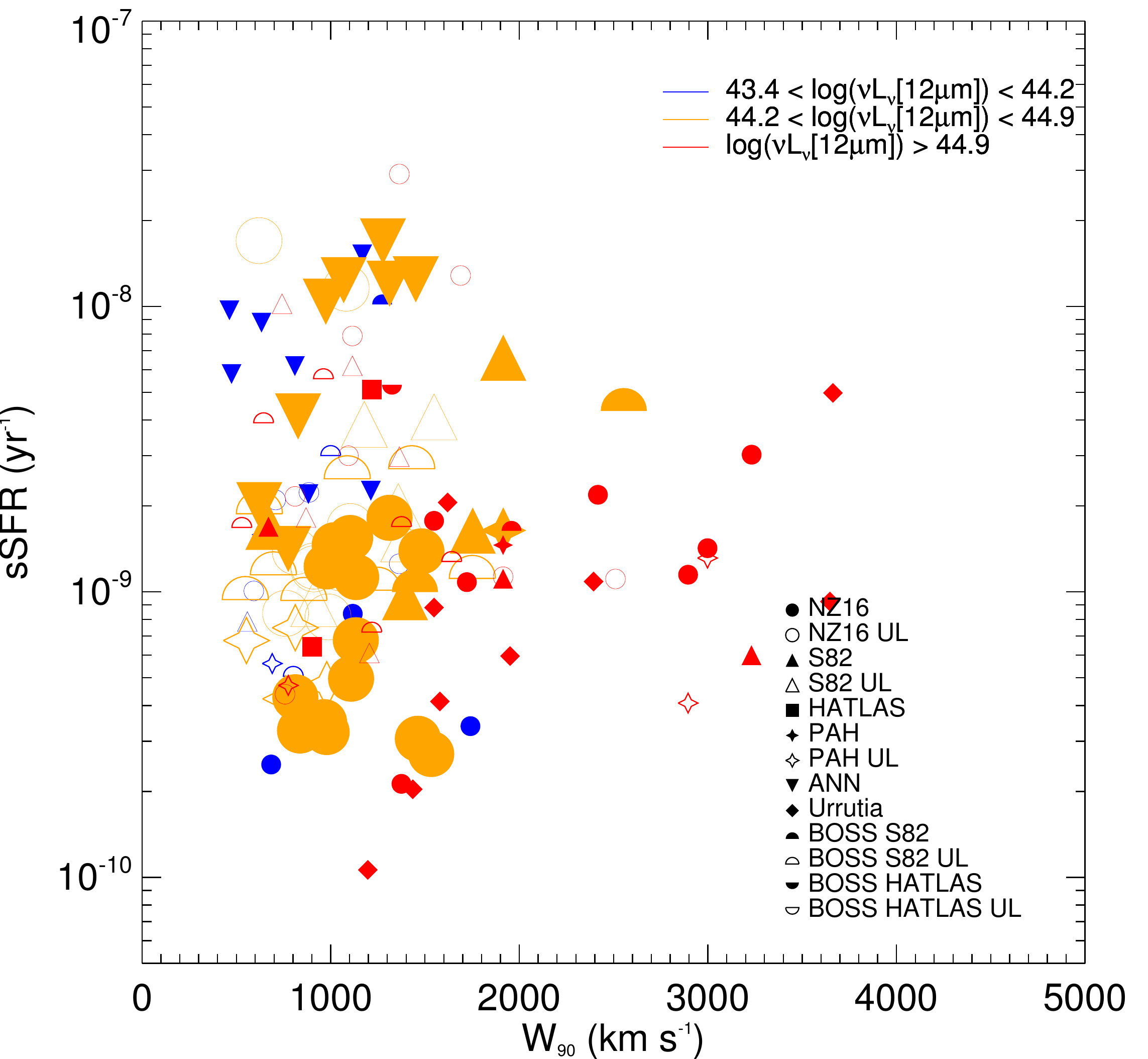}
\includegraphics[trim =  5.2cm 1cm 45cm 39.5cm , clip = true, scale = 0.195]{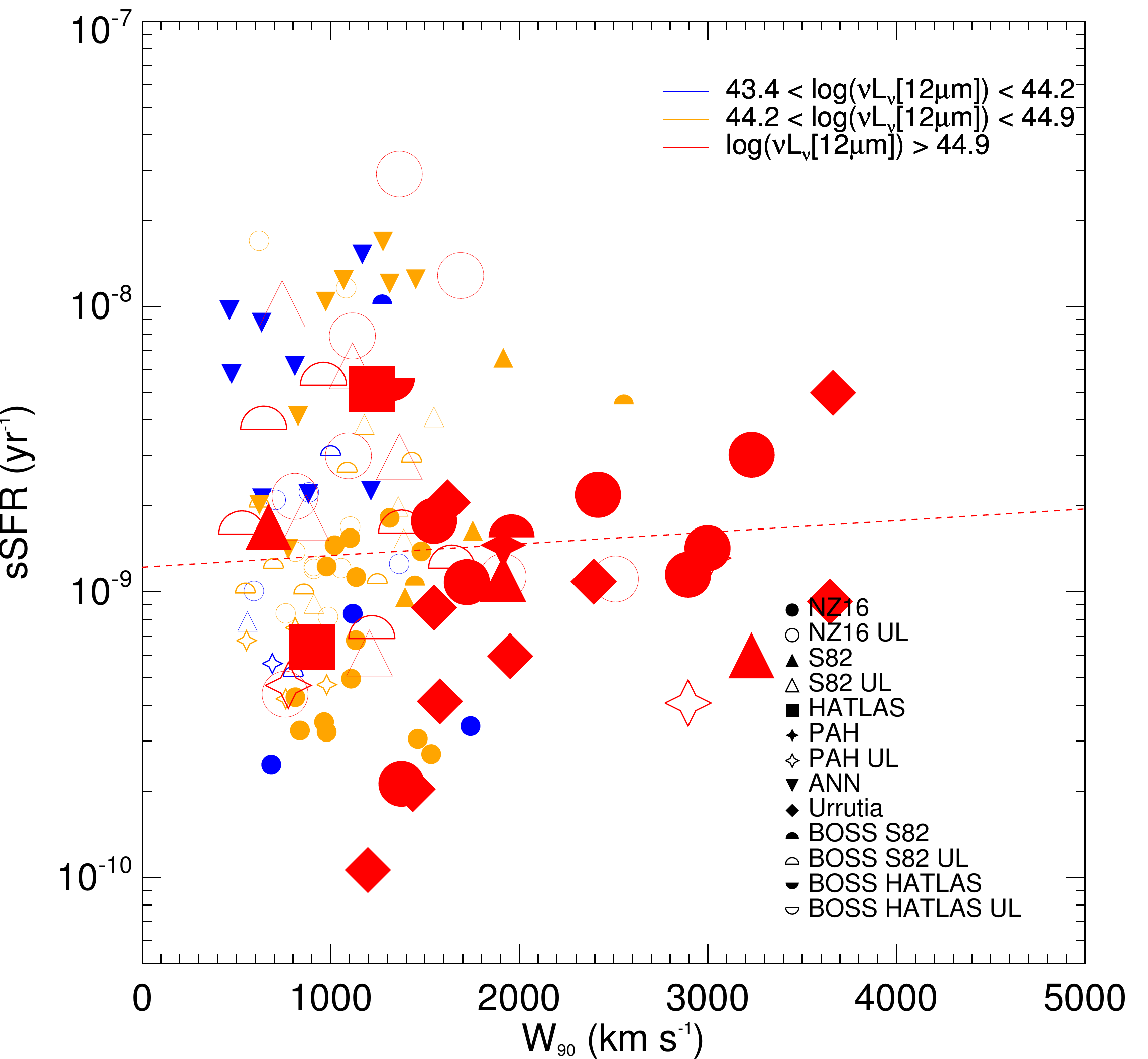}
\caption{Specific star formation rate as a function of [OIII] velocity width, $W_{90}$. The three subpanels are identical, but each highlights the data in a different mid-IR luminosity bin. The dashed line in the highest mid-IR luminosity panel shows a simple linear fit to the data. We do not detect a statistically significant correlation between sSFR and $W_{90}$ in either mid-IR luminosity bin.}
\label{sig_1}
\end{center}
\end{figure}

\begin{figure*}
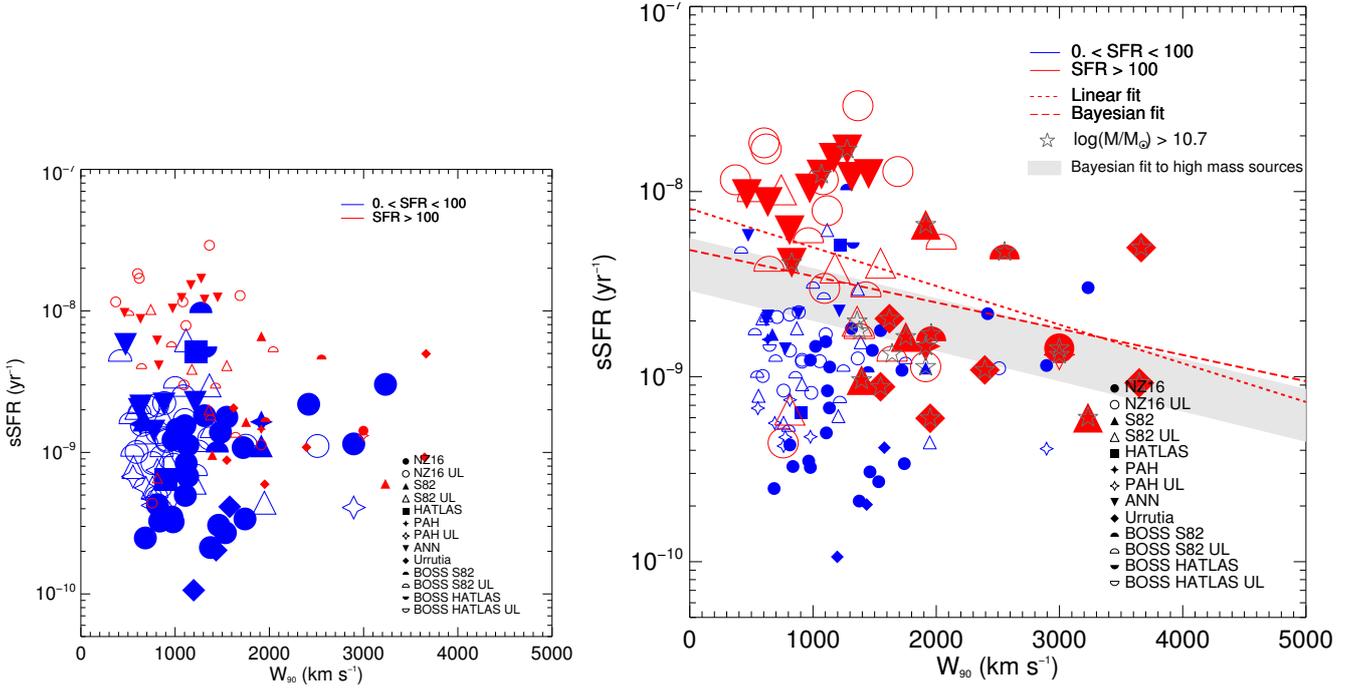

\begin{center}
\includegraphics[trim =  49cm 0cm 1cm 40cm , clip = true, scale = 0.195]{ssfr_project_no2_1_w90-eps-converted-to.pdf}
\includegraphics[trim =  49cm 0cm 1cm 40cm , clip = true, scale = 0.255]{ssfr_project_no2_2_w90-eps-converted-to.pdf}
\caption{Specific star formation rate as a function of [OIII] velocity width, $W_{90}$. The two sub-panels are identical, but each highlights the data in a different SFR bin. The dashed line show a simple linear fit to the data which do not take into account the upper limits on sSFR. We therefore also show the resulting fit using an advanced regression method by \citet{Kelly_2007} which takes into account the upper limit measurements. These fits are shown as long dashed lines. We detect a statistically significant negative correlation between sSFR and $W_{90}$ in the high SFR bin. In the high SFR bin panel, we also highlight sources with $\rm{log}(M_{\rm{stellar}}/M_{\sun}) > 10.7$ and show the resulting fit to the high mass data and its $\pm 1\sigma$ confidence region in the dark gray shaded region.}
\label{sig_2}
\end{center}
\end{figure*}

\section{Quantifying the impact of AGN feedback}

AGN feedback onto the star formation can be negative, if the effects of the AGN lead to suppression of star formation \citep{Cano-Diaz_2012, Lanz_2015, Shimizu_2015, Carniani_2016}, or positive, if the star formation is enhanced \citep{Cresci_2015, Cresci_2015b}. While galaxy formation theory predicts that the net impact should be negative \citep{Croton_2006}, observational evidence for a direct impact of AGN-driven winds on star formation has been scarce and self-conflicting. 

\citet{Farrah_2012}, for example, studied a sample of 31 reddened quasars at $0.8 < z < 1.8$ that were selected based on deep UV absorption troughs which are unambiguous signatures of radiatively driven outflows powered by an AGN. They study the relationships between the velocities of the AGN-driven outflows in these broad absorption line quasars (BALs) and their starburst infrared luminosities. There appears to be a significant anti-correlation between the velocities of the outflows and the starburst contribution to the total infrared luminosity. The authors conclude that this observation shows that the outflows couple to the gas in such a way that star formation can contribute less than 25\% to the overall IR luminosity. \citet{Farrah_2012} note that the observed effect is only relative in the sense that outflows reduce the contribution of star formation to the total infrared luminosity, no direct correlation between outflows and absolute starburst luminosities were found. 

More recently, \citet{Balmaverde_2016} investigated the relationships between optical outflow signatures traced by [OIII]$\lambda\lambda4959,5007$ and [OII]$\lambda\lambda3726,3729$ and star formation rate in a large sample of 224 type-1 quasars selected from the SDSS at $z < 1$. Star formation rates are measured using infra-red luminosities derived from \textit{Herschel} photometry from serendipitous archival coverage. No correlation between the velocity of the outflow and star formation rate is found. SFRs seem to be slightly higher in their subsample with stronger outflow signatures. Based on these results, the authors disfavor the negative feedback scenario and suggest that other mechanisms are possibly responsible for quenching star formation. 

In a more detailed analysis, using integral field observations of a powerful quasar at $z = 2.4$, \citet{Cano-Diaz_2012} show that star formation traced by narrow H${\alpha}$ emission on several kpc is heavily suppressed in regions with the strongest outflows, as traced by the [OIII] velocity shift and dispersion. This leads to a non-uniform spatial distribution of star formation, whose overall SFR is $\sim 100$~M$_{\odot}$/year in this galaxy. 

We are learning from these and many more analyses that quenching of star formation through AGN feedback seems to be a highly complex phenomenon where several different factors might play a role. Suppression of star formation might be highly dependent on the overall stellar mass, size and dust content of the galaxy. The propagation of winds in an individual galaxy will also be highly dependent on the inhomogeneity of the ISM and the distribution of the gas on galaxy-wide scales (i.e. disk vs. bulge-like distribution). Simple correlations between star formation rate and various tracers for AGN luminosity, accretion rate, wind/outflow signatures are hard or even impossible to establish \citep{Page_2012}. 

In this paper, we investigate the impact of AGN-driven outflows on the star formation using specific star formation rate (sSFR), i.e. star formation rate normalized by the stellar mass of the galaxy. We quantify the outflow signatures here by $W_{90}$, the emission line widths where 90\% of the cumulative flux is enclosed. As \citet{Liu_2013b} show, these emission line widths reflect well the typical bulk velocities of the outflowing gas within a numerical factor of about $2-3$, whether measurements such as the velocity offset are more sensitive to projection and orientation effects and are zero in case of perfectly spherically symmetric outflows.

In the following, we use absolute value of $W_{90}$ to quantitatively characterize our results, but use the term `outflow strength' when making relative statements. A source with a higher outflow strength will simply be a source with a higher $W_{90}$ measurement.

\subsection{$W_{90}$ vs. sSFR as a function of total AGN luminosity}
 
We first investigate how the specific star formation rate (sSFR) relates to outflow strength as a function of AGN luminosity. Since most of our sources (except the ones in the Urrutia subsample) are type-2 AGN, where the optical part of their spectral energy distribution is dominated by the light of the host galaxy, the bolometric luminosity of the AGN has to be estimated using indirect measurements. 

In Section 3.2 we have demonstrated that the [OIII] luminosity in type-2 AGN can serve as a good proxy for the total bolometric luminosity of the quasar. But since we are now investigating how sSFR relates to outflow strength (velocity width of [OIII]) in different bins of AGN luminosity, we are seeking an [OIII]-independent measurement for the total bolometric AGN luminosity in order to avoid any biases in our analysis.

The mid-infrared luminosities in type-2 quasars are dominated by the quasar and not by the host galaxy \citep{Lacy_2004, Stern_2005} and are tracers for the total luminosity of the AGN \citep{Richards_2006}. The \textit{Wide Infrared Survey Explorer} \citep[WISE; ][]{Wright_2010} has imaged the whole sky at $3.4, 4.6, 12$ and $22\mu$m and provides band-matched all sky catalogs. All sources in this paper are detected by WISE and we calculate $K-$corrected monochromatic luminosities at rest-frame $12\mu$m. To do so, we interpolate between the rest-frame $12$ and $22\mu$m fluxes using a power law to compute the rest-frame $\nu L_{\nu}$ at $12\mu$m. The sources in the total sample span infrared luminosities $43.4 < \rm{log}(\nu L_{\nu, 12 \mu m}/\rm{(erg/s)}) < 46.7$ with a mean luminosity of log$(\nu L_{\nu, 12 \mu m}/\rm{(erg/s)}) = 44.7$. These mid-infrared based estimates for the total AGN luminosity are in excellent agreement with the [OIII] based estimates, with a median ratio between the two of $0.95\pm 0.01$. This ratio is not expected to be unity, as no bolometric correction has been applied to the mid-IR luminosities. Since we are are utilizing $\nu L_{\nu}$ at $12\mu$m as a relative estimator for total AGN luminosity, we abstain here from applying a bolometric correction.

We then divide up our galaxies into three groups of mid-IR luminosity: one low luminosity regime $43.4 < \rm{log}(\nu L_{\nu, 12 \mu m}/\rm{(erg/s)}) < 44.2$, one high luminosity regime $\rm{log}(\nu L_{\nu, 12 \mu m}/\rm{(erg/s)}) > 44. 9$ and one intermediate luminosity regimes ($44.2 < \rm{log}(\nu L_{\nu, 12 \mu m}/\rm{(erg/s)}) < 44.9$). Figure \ref{sig_1} shows the sSFR as a function of $W_{90}$, color coded by L$_{\nu}$ at $12\mu$m. The three subpanels show the same data, each of them highlighting a different luminosity regime using bolder symbols. 

We first perform a Spearman rank correlation test for the data in each luminosity bin and do not detect a significant correlation between the outflow strength $W_{90}$ and sSFR in any of the luminosity regimes. A Spearman rank test results only in a probability of $50-70$\% of a correlation. Additionally, we perform a Spearman rank correlation test in each luminosity bin but this time only taking into account sources with $\rm{log}(M_{\rm{stellar}}/M_{\sun}) > 10.7$. These measurements are least affected by the limiting magnitude $K = 18.3$ of the UKIDSS survey which otherwise might introduce a redshift-dependent lower mass limit on our data. We elaborate on this possible bias in more detail in the next subsection, but note here that no correlation arises even if only high mass sources are considered. 
 
The lack of a correlation between $W_{90}$ and sSFR for AGN selected on their mid-IR luminosity shows AGN power is not the only and probably also not the most important driver in regulating sSFR in AGN host galaxies. In the next Section we therefore investigate how the host galaxies' gas content impacts sSFRs.   

\subsection{$W_{90}$ vs. sSFR as a function of star formation rate}

We next investigate the relation between $W_{90}$ and the sSFR as a function of SFR of the galaxies. This is particularly interesting since SFR traces the absolute gas mass and is highest in the most gas-rich systems. We study the correlation between $W_{90}$ and sSFR in two bins of SFR: $0-100$~M$_{\odot}$/year and SFR > $100$~M$_{\odot}$/year. Similar to Figure \ref{sig_1}, Figure \ref{sig_2} shows $W_{90}$ as a function of sSFR, with data points in the SFR bins highlighted in different subpanels, respectively. While no significant correlation between $W_{90}$ and sSFR is detected at low SFRs, it is remarkable that at high SFR, $W_{90}$ significantly anti-correlates with sSFR. At higher outflow strengths, i.e. at higher velocity widths of the [OIII] emission line, the specific star formation rate seems to be heavily suppressed. A Spearman rank correlation test confirms a significant correlation with 99.9\% probability.

We furthermore perform a linear regression on the data in the highest luminosity regime and show the resulting fit with a dashed line. \citet{Kelly_2007} has developed a Bayesian method to account for measurement errors and upper limits in linear regression of astronomical data. This method assumes that measurement errors are Gaussian with zero mean and that the intrinsic scatter of the dependent variable (sSFR in our case) is Gaussian around the regression line. IDL routines to use this advanced linear regression are available through the \textit{astrolib}. We show the result of that linear regression, taking properly into account the upper limits on sSFR, in Figure \ref{sig_1} with the long dashed line. 

Part of the observed correlation could be driven by the fact that the sources from the ANN subsample which are preferentially populating the low-$W_{90}$/high-sSFR area of the parameter space are on average of lower redshift than the sources populating the high-$W_{90}$/low-sSFR area of the parameter space. As we have discussed in Section 3.3, the evolution of the $K-$band magnitude as a function of redshift \citep{Kauffmann_1998} introduces a redshift-dependent lower limit on stellar mass. This in turn introduces a redshift-dependent upper limit on sSFR. Taking the extreme ends of the redshift distribution of the sources in the highest bin of SFR (bold red data points in Figure \ref{sig_2}), $z_{\rm{min}} \sim 0.2$ and $z_{\rm{max}} \sim 0.8$, we estimate that the redshift-dependent lower limit on stellar mass could introduce a decrease of sSFR as a function of $W_{90}$ of maximum 0.8~dex. 

We therefore highlight the sources with stellar masses $\rm{log}(M_{\rm{stellar}}/M_{\sun}) > 10.7$ in Figure \ref{sig_2}. At the $K-$band depth of the UKIDSS survey, $K = 18.3$, this is the lower limit on stellar mass for a $z = 1$ galaxy \citep{Kauffmann_1998}. Despite low-number statistics, a Spearman rank correlation test confirms a correlation with 99\% probability in the high SFR bin. We then follow the same fitting technique discussed above, taking into account the upper limits on the data, and show the resulting fit and its $\pm 1\sigma$ confidence region in gray. The confidence region of the fit to the high-mass data is in excellent agreement with the fits to the data where no mass cut has been employed. This test confirms that the strong observed correlation between sSFR and $W_{90}$ in the highest bin of SFR is not driven by the redshift-dependent lower limit on stellar mass in the UKIDSS survey.

Another potential bias in this correlation could arise through a redshift dependence of the sSFR. The normalization of SFR main sequence increases over the redshift range $z=0$ to $z=1$ such that galaxies at higher redshift on average have higher sSFR than galaxies at lower redshift. We therefore normalize the sSFRs in Figure \ref{sig_2} by the average sSFR at the corresponding redshift of the source using the SFR-main sequence evolution from \citet{Elbaz_2011}. Within the fitting uncertainties, we recover the same negative correlation between normalized sSFR and $W_{90}$ at high SFRs showing that a potential redshift dependence of sSFR is not impacting the results of this analysis.

\begin{figure}
\begin{center}
\includegraphics[trim =  3cm 0cm 45cm 40cm , clip = true, scale = 0.2]{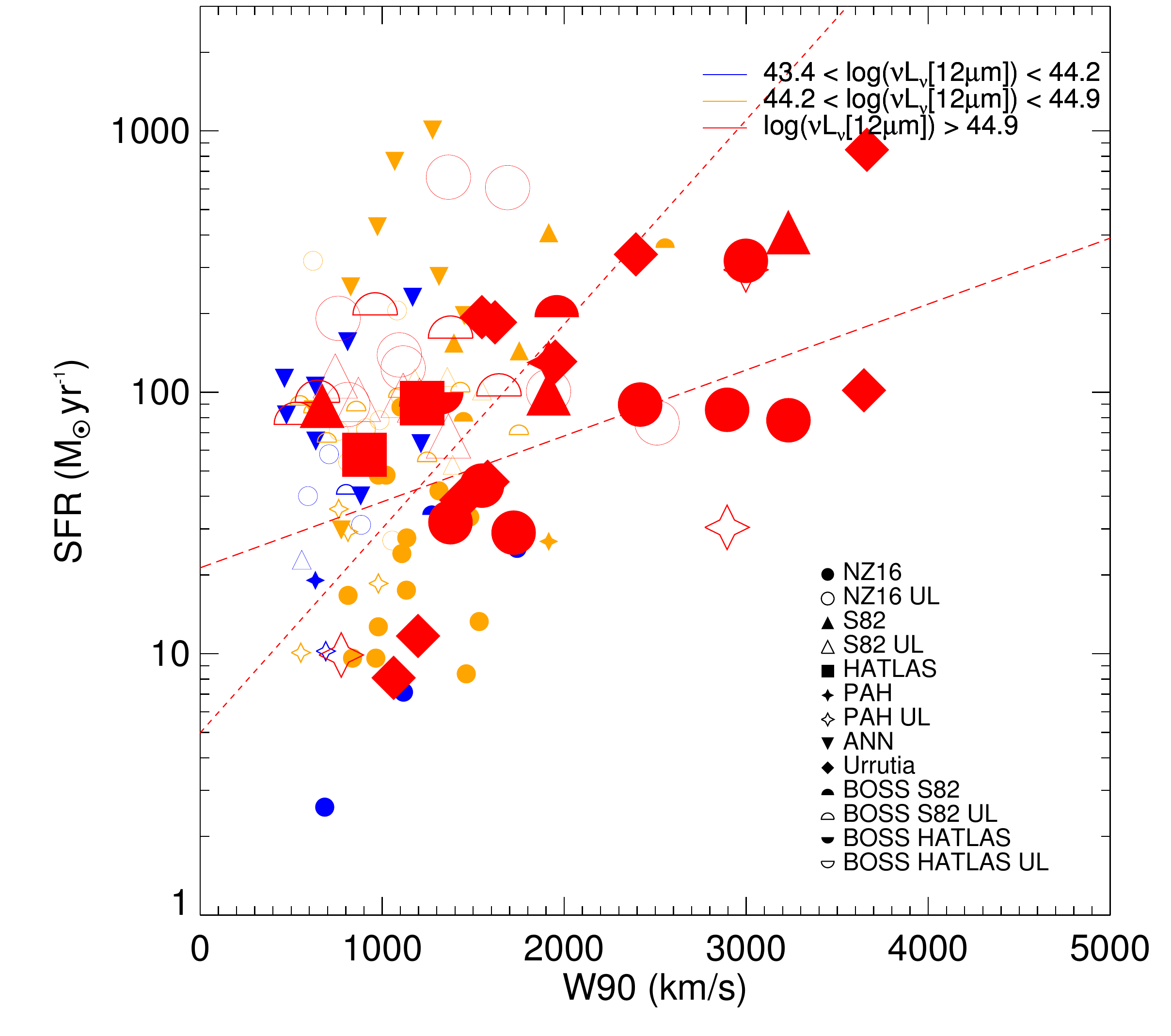}
\caption{Star formation rate as a function of [OIII] velocity width, $W_{90}$, where differently colored data points correspond to different bins of IR-luminosity. We highlight the data in the bin of highest AGN IR luminosity and show that SFR (a measure for the galaxies' gas content) correlates strongly with $W_{90}$. Selecting AGN based on their luminosity results in a large range of initial galaxy properties, specifically as gas content.} 
\label{sfr_w90}
\end{center}
\end{figure}

\begin{figure}
\begin{center}
\includegraphics[trim= 5cm 1cm 2cm 1cm clip = true, scale = 0.3]{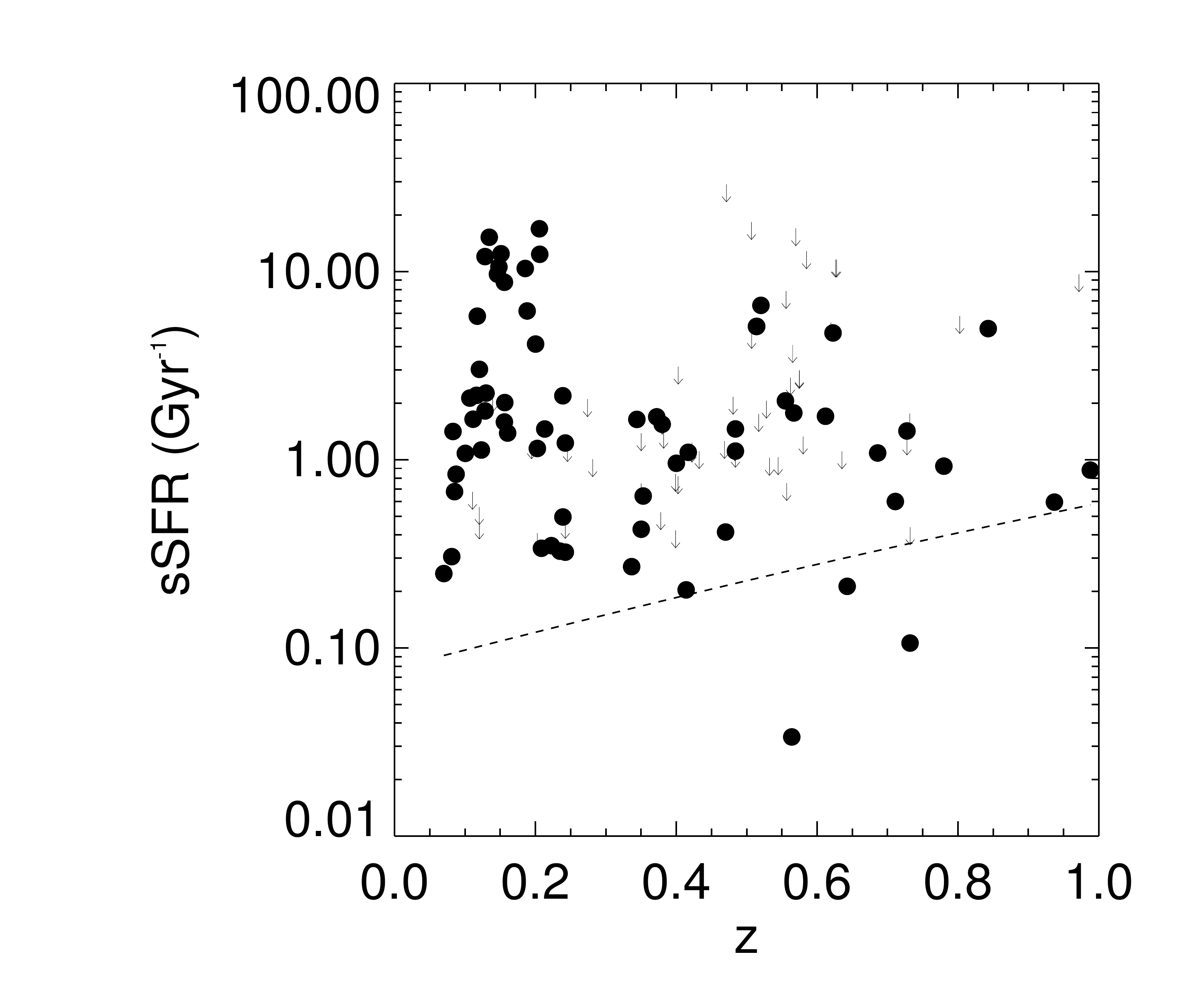}
\caption{Evolution of the specific star formation with redshift. The dashed line shows the evolution of the main sequence adapted from \citet{Elbaz_2011} \citep[see also][]{Balmaverde_2016}. The sources in our sample are on average about 1~dex above the main sequence.}
\label{ms}
\end{center}
\end{figure}

\section{Discussion}

\subsection{Is AGN luminosity related to galaxy growth or quenching?}

In the previous Section we have shown that outflow strength, i.e. the velocity width of the [OIII] emission line, does not correlate with quenching or enhancement of the specific star formation rate (negative vs. positive feedback) when the entire sample is considered as a whole. As speculated above, the situation seems to be far more complex and the relationships between AGN-driven winds and specific star formation rates of the hosts appear only when we examine the multi-dimensional parameter space that includes AGN luminosities, outflow velocities, SFRs and sSFRs. 

As seen in Figure \ref{sig_1}, a high luminosity of the AGN (which we estimate here using the monochromatic rest-frame $12\mu$m luminosity) is a prerequisite for having a high-velocity outflow. This is consistent with observations by \citet{Zakamska_2014}. Studying a large sample of type-2 AGN selected from the SDSS, they show that AGN luminosity and [OIII] line width strongly trace each other, albeit with significant scatter. The median [OIII] velocity widths in our three $\nu$L$_{\nu}$ bins are 850, 1060 and 1900 ~km~s$^{-1}$, respectively, consistent with the findings in \citet{Zakamska_2014}. The most extreme examples of quasar-driven outflows at the peak of quasar activity, $z \sim 2$, would be found at the extreme end of this correlation, with outflow velocities $W_{80} \sim 2000-5000$~km~s$^{-1}$ \citep{Brusa_2015a, Zakamska_2016b}. Theoretical \citep{Zubovas_2012} and observational arguments \citep{Veilleux_2013, Zakamska_2014} suggest that there might be a threshold in AGN luminosity $L_{\rm{bol}} \sim 3 \times 10^{45}$~erg~s$^{-1}$, above which the AGN is powerful enough to power winds that are fast enough to escape the galaxy potential. The galaxies in our sample are all above this threshold, and not surprisingly the median [OIII] line widths are $W_{90} \sim 1100$~km~s$^{-1}$, a velocity typically too large to be explained by fast rotating or disturbed gas disks \citep{AlaghbandZadeh_2012}. In the highest $\nu$L$_{\nu}$ regime, only four objects are below $W_{90} \sim 1100$~km~s$^{-1}$ (Figure \ref{sig_1}). The broad range of [OIII] line widths ($\sim 1000 - 4000$~km~s$^{-1}$) in the highest $\nu$L$_{\nu}$ bin is not surprising, given the scatter in the $\nu$L$_{\nu}$-$W_{90}$ correlation seen in \citet{Zakamska_2014}. The fact that we observe the highest [OIII] velocity widths in the highest bin of $\nu$L$_{\nu}$ (see Figure \ref{sig_1}) is therefore in agreement with the assumption that more powerful AGN are more capable of powering high-velocity winds.

Although AGN luminosity is important for driving high velocity outflows, by itself it does not appear to be correlated with SFR or sSFR. In Figure \ref{sfr_w90} we use the same color-coding for the different $\nu$L$_{\nu}$ bins, but this time show the absolute SFR (as opposed to the sSFR in Figure \ref{sig_2}) as a function of $W_{90}$. In each of the AGN luminosity bins the range of observed SFRs is similar and very broad with star formation rates ranging between 10 to 1000~M$_{\odot}$/year. The lack of a relationship between AGN luminosity and SFRs in our sample is consistent with findings in other studies. For example, \citet{Stanley_2015} find, for a sample of 2000 X-ray selected AGN at $0.2<z<2.5$, that star formation rate in AGN hosts is not strongly correlated with AGN luminosity. This lack of a relationship is somewhat surprising given that the overall cosmic evolution of star formation and black hole growth show very similar histories \citep{Boyle_1998}. \citet{Stanley_2015} conclude that short-time variability of accretion rates can wash out the long-term relationship between SFRs and L$_{\rm AGN}$ in a sample selected based on AGN luminosity. 

Additionally, the bold red data points and (long) dashed lines highlight the data in the highest luminosity bin and show the fits to the data, respectively. SFR correlates significantly with $W_{90}$ in high-luminosity AGN. Assuming that star formation rate is a tracer for the galaxies' gas content \citep{Kennicutt_1998, Doyle_2006, Leroy_2011}, this observation means that, similarly to the lower AGN-luminosity bins, high-luminosity AGN trace galaxies with a wide range of gas contents. As discussed above outflow velocity is dependent on AGN luminosity and only in the most powerful AGN winds can be accelerated to high velocities. The positive correlation between SFR and $W_{90}$ now additionally shows that not only is AGN luminosity critical for wind acceleration but so is gas content. The highest velocity winds are therefore found in the most powerful AGN hosted by the most gas-rich galaxies.


Several other authors, looking for the smoking gun of negative AGN feedback, who were initially trying to establish a negative correlation between SFR and $L_{\rm{AGN}}$, have found the opposite \citep{Balmaverde_2016}. By just investigating the relation between SFR and outflow strength, we find very similar results in this paper (see Figure \ref{sfr_w90}). But an AGN selection based on $L_{\rm{AGN}}$ results in a selection of a wide range of initial galaxy properties and different gas-wind coupling, thus we advocate the use of specific star formation rate instead, which measures the relative effect of AGN feedback and can isolate the effect of AGN feedback on galaxies with similar host galaxy properties.

\subsection{How do SFR and feedback signatures relate to AGN type?}

In Section 4.2 we have shown that sSFR significantly decreases as a function of $W_{90}$ for galaxies selected based on their high star formation rates (SFR > 100~M$_{\odot}$/year). Using $\sim 200$ type-1 quasars at $z < 1$ selected from the SDSS, \citet{Balmaverde_2016} performed a similar analysis. They find no such correlation, so we investigate the difference between our studies in this section.

In Figure \ref{ms} we compare the sSFR of the galaxies in our sample to the main sequence of SF \citep[adopted from ][]{Elbaz_2011}. Interestingly, except for a handful of exceptions, most of the sources in our sample lie significantly above that main sequence, meaning that with respect to the overall population, they are highly star forming. This is especially striking given that luminous quasars in our sample tend to occupy ellipticals, not disks \citep{Wylezalek_2016}.

Despite quite uncertain host galaxy mass measurements (which is an extremely challenging quantity to measure in type-1 quasars), \citet{Balmaverde_2016} show that the sources in their sample broadly follow the main sequence of star formation with a relatively large scatter of about 1~dex. Relative to the main sequence these sources seem to be already more `quenched' compared to the sources in our sample. This might be surprising since both samples have been selected on similar properties such as [OIII] luminosities and redshift range. This large offset between the \citet{Balmaverde_2016} sample and ours in terms of absolute values of sSFR can partly be explained by different conversion factors from $L_{\rm{IR}}$ to SFR. This would only account for a shift of 0.2 dex towards lower sSFR for our sources. 

Therefore, the more likely explanation for the difference between \citet{Balmaverde_2016} results and ours is that the two samples are tracing different stages of the AGN evolution. Contrary to the standard AGN unification model where type-1 and type-2 AGN are intrinsically the same objects observed under different angles, the evolutionary unification model suggests that type-2 AGN are more likely to represent an earlier, dust-enshrouded phase of the AGN than type-1. This happens as the merger-triggered, dust-obscured black holes (type-2 AGN) become more powerful, clear out the environment and become type-1 AGN \citep[e.g.][]{Sanders_1988, Hopkins_2006, Menci_2008, Somerville_2008, Glikman_2015}. There is some observational evidence that type-2 quasars are the most star-forming among all quasar types \citep{Shi_2007}. Therefore, type-1 AGN might preferably be found in more evolved, `quenched', galaxies. This possibility also finds support in earlier work using observations from the \textit{Spitzer Space Telescope} \citep{Shi_2007, Zakamska_2008}. 

Recently, \citet{Xu_2015b} compared sSFR between two samples of $24\mu$m selected type-1 and type-2 AGN at $z <1$ in the Local Cluster Substructure Survey \citep{Smith_2010}. Although they state that their type-1 and type-2 AGN have sSFR that are consistent with the main sequence, a closer inspection of the data at $0.3 < z < 1$ shows that their type-2 AGN are consistently found above the main sequence while their matched type-1 counterparts tend to lie below the main sequence. The AGN in \citet{Xu_2015b} tend to be of slightly lower AGN luminosity \citep{Xu_2015a} than the AGN in \citet{Balmaverde_2016} and in this work. There might therefore be tantalizing evidence that type-1 AGN preferentially reside in host galaxies with lower levels of sSFR, possibly connected to them being at a later evolutionary stage than type-2s. We abstain here from drawing quantitative conclusions given the uncertainties in the stellar mass derivation for type-1 AGN. 

\subsection{Do optical outflow signatures trace star formation rate quenching?}

Turning back to the impressively strong correlation between sSFR and $W_{90}$ for SFR > 100 ~M$_{\odot}$/year, allows us to speculate whether this might be a signature of negative AGN feedback. As just discussed, most galaxies in our sample lie significantly above the main sequence of star formation (Figure \ref{ms}) and the strong anti-correlation between sSFR and $W_{90}$ is found only in the highest bin of star formation. This leads us to suggest that the cleanest examples of AGN feedback suppression of star formation are to be found in the most gas-rich systems. There, the wind powered by the AGN presumably can couple best to the available gas in the galaxy. This demonstrates that signatures of the relative impact of AGN feedback, i.e. the amount of SFR normalized by stellar mass, are better observables than absolute measurements. 

Since we select galaxies based on their SFR (i.e. their gas content), the negative correlation between sSFR and $W_{90}$ is primarily driven by an increase in stellar mass. Galaxies in which strong outflow signatures are found are therefore more `quenched' with respect to their size and stellar mass than galaxies with weaker outflow signatures, an intriguing result consistent with AGN accelerating the evolution of a galaxy and having a `negative' impact feedback on their star formation history. Additionally, galaxies with high $W_{90}$ and low sSFR are on average more IR-luminous than galaxies with low $W_{90}$ and high sSFR. For gas-rich galaxies (SFR > 100~M$_{\odot}$/year) where coupling between gas and wind seems to reach its maximum, higher luminosity AGN therefore power higher velocity outflows and these galaxies are found to have suppressed sSFR. Overall, the distribution of total AGN luminosities in the subsample of type-2 AGN with SFR$>100$~M$_{\odot}$/year is identical to the distribution of the total sample.

There is a notable difference in the $W_{90}$ distribution of the high SFR subsample compared to the total type-2 AGN sample. While the overall distribution of outflow strengths of the total sample peaks at $W_{90} \sim 1100$~km/s (see Figure \ref{oiii_histo}) with a long tail towards more extreme outflow strengths of $2000-4000$~km/s, the distribution of outflow strengths in the high SFR (SFR$~> 100$~M$_{\odot}$/year) subsample is vastly different. The median outflow velocity is significantly higher ($W_{90} \sim 1400$~km/s) and the overall distribution is close to flat and includes the highest velocity objects with $W_{90} \sim 4000$~km/s. Only about 25\% of the high SFR objects have outflow strengths $W_{90} < 1000$~km/s. Since these sources are selected based on their high SFRs, potential contribution of starburst-driven winds has to be explored.

The very high values of $W_{90}$ in the key high-SFR subsample shown in Figure \ref{sig_2} (right panel) strongly suggest that the outflows are not star-formation driven. Indeed, several observational studies suggested that the ionized gas velocities can be used as a discriminant between SF-driven and AGN-driven winds, with AGN-driven winds showing significantly higher velocities \citep{Rupke_2005, Rupke_2013, Hill_2014}. Furthermore, numerical simulations also suggest that AGN-driven outflows proceed faster (though for a shorter period of time) than the SF-driven ones \citep{Narayanan_2008}. 

To explore quantitatively whether the $W_{90}$ values we see in our sample can arise in SF-driven outflows, we use the sample of \citet{Hill_2014}. The authors analyzed SDSS spectra of 36 ultra-luminous infrared galaxies (ULIRGs) selected from the \textit{Infrared Astronomical Sattelite} (IRAS) 1~Jy sample and measured outflow strengths in an identical manner as we do, using the $W_{90}$ parameter of the [OIII] emission line. The complete IRAS 1~Jy sample was originally presented by \citet{Kim_1998}. \citet{Kim_1998} also estimated total IR luminosities for the sources using a calibration based on the sum of the flux in all IRAS bands. We now derive SFRs based on the total IR luminosities presented in \citet{Kim_1998} following \citet{Kennicutt_1998}. We further exclude sources with an equivalent width (EW) of the 6.2~$\mu$m PAH feature EW(PAH[6.2~$\mu$m])$<0.3~\mu$m which excludes sources where contribution from an AGN to the total luminosity exceeds 50\% \citep{Armus_2007} resulting in a final non-AGN ULIRG sample size of 26 sources.

The median SFR of the non-AGN ULIRG sample is $\sim 250$~M$_{\odot}$/year and the median outflow strength is $W_{90} \sim 670$~km/s. The median SFR of our high-SFR subsample is $\sim 200~$M$_{\odot}$/year, very similar to the ULIRG sample. However, the median outflow strength in our sample is $W_{90} \sim 1400$~km/s, more than twice as high as in the ULIRG sample. Comparing these two galaxy samples that are so similar in terms of total SFRs but have vastly different  ionized gas kinematic properties reassures that the outflows in our objects are, despite high star formation rates, most likely not dominated by star formation-driven winds. Interestingly, when we compute the median $W_{90}$ for IRAS sources with EW(PAH[6.2~$\mu$m])$>0.3~\mu$m (i.e. their bolometric luminosities are AGN-dominated), we find a median $W_{90} = 1400$~km/s, an almost identical median outflow strength as in our type-2 high-SFR subsample. This shows that only in AGN-dominated sources such high outflow velocities can be reached and these outflows are almost certainly powered by the AGN.

For the typical wind velocities of the sources studied in this paper, the travel time of the gas clouds can be estimated to be $\sim 10^7$~yr to reach the outskirts of their host galaxy at $\sim 10$~kpc distance. Typical quasar lifetimes have been estimated through various methods in the past to be of the order of a few $10^7-10^8$~yr \citep{Haehnelt_1993}. Since  in this work we are observing the effects of sSFR quenching while the AGN is still active, we estimate that sSFR quenching as observed here happens on timescales $< 9\times10^7$yr. This timescale is not too different from quenching timescales estimated in simulations \citep{Zubovas_2012}.

The second possibility is that the anti-correlation between $W_{90}$ and sSFR is due to the dynamics of the wind. We already noted that the strong negative correlation between $W_{90}$ and sSFR is driven by the increase of stellar mass as a function of $W_{90}$ rather than by the decrease of SFR as a function of $W_{90}$. At fixed SFR, however, such a correlation could be expected simply due to the fact that with increasing stellar mass the winds will have to overcome a higher galactic potential and would thus expected to be of higher velocity. We therefore investigate now if the observed correlation is consistent with models for outflows that propagating through different galaxy potentials. \citet{Zubovas_2012b}  \citep[for a review, see also][]{King_2015} show that the expected outflow velocity of an energy-driven outflow in an isothermal potential would be 

\begin{equation}
v_e \simeq 925 \left( l \frac{f_c}{f_g}\right) ^{1/3}\sigma^{2/3}\rm{km~s^{-1}},
\label{eq:2}
\end{equation}
where $v_e$ is the outflow velocity, $\sigma$ the stellar velocity dispersion, $f_c$ the gas fraction of the galaxy at the epoch when the $M_{\rm{BH}}-\sigma_{*}$ relation was established, $f_g$ the current gas fraction through which the wind propagates and $l$ a scaling factor to allow for deviations from the Eddington luminosity. The fluctuations $l$ have almost no effect on the outflow \citep{King_2015} and $f_c$ can be neglected since it serves as a proxy for the normalization of the $M_{\rm{BH}}-\sigma_{*}$ relation. We can estimate the stellar velocity dispersion $\sigma_{*}$ from the galaxy stellar masses using the Faber-Jackson relation $L_{*} \propto \sigma_{*}^3$ and assuming a constant mass-to-light ratio, such that $M_{\rm{stellar}} \propto \sigma_{*}^3$. Combining this with equation \ref{eq:2} and taking $f_g = \frac{M_g}{M_{\rm{stellar}}}$ leads to 

\begin{equation}
\frac{1}{M_{\rm{stellar}}} \propto v_e^{-9/5} M_g^{1/3}.
\end{equation}

At fixed SFR, $M_g$ can be considered a constant. This is a valid assumption for our analysis given that we investigate feedback signatures in fixed bins of SFR. Then, sSFR is expected follow the relation $sSFR \propto v_e^{-9/5}$. This relation is steeper than the one we observe in Figure \ref{sig_2}. The \citet{King_2015} models predict winds with only half to a third the velocity compared to the observed value, when we normalize the two relations at $W_{90} = 1000$~km~s$^{-1}$. The observed wind velocities thus exceed the model velocities.

We note that these models are based on simple assumptions, i.e. the gravitating density and the ISM density distributions are assumed to be isothermal and the wind is assumed to be purely energy-driven. Therefore, we do not not exclude the possibility that the observed negative correlation between sSFR and $W_{90}$ is a result of modification of wind dynamics due to different stellar masses, i.e. gravitational potentials.

\subsection{Is the far-IR a good measure for varying (s)SFR?}

A possible caveat of our observations concerns the use of the far-IR emission as a measure of instantaneous star formation. The far-emission is considered to be a good estimator for SFR because no correction for dust attenuation has to be performed like for other optical SFR indicators. Especially in highly obscured galaxies, such as starburst galaxies, essentially all the light emitted in the UV is absorbed by dust and re-emitted in the IR. Precisely for such galaxies, the far-IR is the ideal tracer for SFR. As shown in Figure \ref{ms} and by \citet{Wylezalek_2016}, luminous type-2 AGN tend to indeed reside in very dusty, highly star-forming galaxies. However, the far-IR, like all other SFR indicators, is sensitive to the initial mass function. The far-IR is both sensitive to the high mass and low mass end of the IMF since the dust is heated by young stars with a wide range of masses. Thus the IR luminosity may not trace the instantaneous SFR if it varies on timescales $< 100$~Myr \citep{Hayward_2014}.  

It is unclear how much of the IR radiation is contributed by the general stellar radiation field. Using mock SEDs calculated from three-dimensional hydrodynamical simulations of isolated disc galaxies and galaxy mergers, \citet{Hayward_2014} recently showed that while SFRs estimated from the far-IR agree well with true instantaneous SFR of the simulated galaxies, the situation might be different in galaxies undergoing major mergers. There, the far-IR derived SFR may severely overestimate the true instantaneous SFR due to non-negligible contributions from stars formed during or prior to merger coalescence. The magnitude of the overestimate seems to be greater for lower sSFR values. While \citet{Wylezalek_2016} show that a significant fraction of luminous type-2 AGN seem to have undergone a recent merger, it seems that $z<1$ quasar activity is associated with minor, rather than major, mergers. Additionally, the levels of sSFR in this current paper exceed by far the range in sSFR investigated in \citet{Hayward_2014}. Therefore, the effects they discuss are likely less important for our objects, so we can use far-IR luminosities as tracers of star formation.

\section{Conclusions}

In this paper, we collect a wide range of optical to far-IR photometry and optical and near-IR spectra for a sample of 122 type-2 and 10 type-1 AGN at $0.1 < z < 1$. The purpose of this archival data mining is to derive reliable stellar masses, star formation rates and narrow-line kinematics for a representative sample of powerful, intermediate redshift quasars in order to test for observational evidence of quasar feedback. The focus in this paper is to study the relative impact of AGN on their host galaxies, by studying the relationship between specific star formation rate and outflow strength of the quasar-driven wind, parameterized through the velocity width $W_{90}$ of the [OIII] emission line. 

Stellar masses are derived by fitting the SEDs of the type-2 AGN in the optical through near-IR, or by appropriate scaling using host galaxy luminosities for the type-1 AGN. Star formation rates are derived through a number of methods. We primarily use far-IR luminosities either from space-based observatories such as \textit{Herschel} and \textit{Spitzer} or from predictions from an artificial neural network. We also include sources with SFRs estimated from PAH signatures. SDSS spectra are used to measure the kinematics of the [OIII] emission line.

We investigate the relation between sSFR and $W_{90}$ in different bins of rest-frame $12\mu$m mid-IR luminosity $\nu$L$_{\nu}$, a tracer for the total bolometric luminosity of the AGN, and different bins of SFR, a tracer for the gas content of the host galaxies. 

We find that the mean outflow strength, $W_{90}$, is correlated with AGN IR-luminosity and that the strongest outflows are only found in galaxies in the highest bin of AGN IR-luminosity. This observation is consistent with what has been found in other studies \citep{Zakamska_2016}. Only the highest luminosity AGN are capable of driving the most powerful outflows such that there seems to be a lower threshold of AGN luminosity that is needed to power outflows that can overcome the galaxy potential and drive gas outside the galaxy. Almost all AGN in our work are above that derived threshold of $\sim 3\times10^{45}$~erg/s. We show that selecting AGN based on their luminosity results in a wide range of initial galaxy properties, specifically galaxy gas content. While sSFR does not correlate with $W_{90}$, SFR does. The positive correlation of SFR with $W_{90}$ likely reflects on the amount of gas available in the galaxy and is therefore likely a reflection of the coupling strength of the wind to the gas. If AGN feedback is dependent on the coupling between gas and wind then this effect has to be accounted for. The most straightforward way is to select on high gas fractions, i.e. high SFR, where maximal coupling is likely. The best wind-gas coupling is expected in galaxies in which the gas is well spatially distributed. The host galaxies of the sources in our sample are mostly gas-rich elliptical galaxies fueled by minor mergers \citep{Wylezalek_2016} which helps to achieve maximum coupling. 

We therefore further investigate the relation between sSFR and $W_{90}$ where galaxies have been selected based on their SFR, i.e. gas content, to test the relative impact of AGN feedback on the host galaxies. We observe a strong negative correlation between sSFR and $W_{90}$ for galaxies with SFR~$> 100$~M$_{\sun}$~yr$^{-1}$. In galaxies with large SFR the AGN-powered wind can presumably couple best to the available gas. Almost all galaxies in our sample lie significantly above the main sequence of star formation and can be classified as star-forming galaxies. The observed correlation is primarily driven by the increase in stellar mass with $W_{90}$ and shows that galaxies with strong outflow signatures and selected on their gas content are more `quenched' with respect to their size and stellar mass than galaxies with weaker outflow signatures. Despite the galaxies' high SFRs, we have demonstrated that the outflow velocities are too high to  be star-formation driven. Only AGN can drive such powerful outflows. This observation is consistent with the AGN having a `negative' impact through feedback on the galaxies' star formation history.

Alternatively, the negative correlation between sSFR and $W_{90}$ could also be a result of different wind dynamics in galaxies with different stellar masses, i.e. galaxy potentials. Comparison with simple models \citep[e.g.][]{Zubovas_2012b} show that the model wind velocities are lower than the observed ones which would favor the AGN feedback-scenario. However, due to simplistic assumptions in the models and the large scatter in the data, this dynamical scenario cannot fully be excluded. 

The negative feedback signatures we may be observing in this analysis is not the kind of feedback necessary in shaping the present-day luminosity function of elliptical galaxies \citep{Silk_2013} which was likely established at the peak of galaxy formation \citep{Madau_1998} which coincides with the peak of quasar activity at $z \sim 2-3$ \citep{Schmidt_1995}. Our observations are of quasars at $z<1$, which are hosted by galaxies whose stellar component is already well established. The AGN in our sample represent the tail-end of quasar activity and are not going to dramatically alter the stellar mass growth of the universe. But these data allow us to probe the physics of the same process which presumably occurs at $z=2-3$ despite there potentially being some differences due to overall higher gas fractions, SFRs and lower host galaxy stellar masses at high redshift. The great advantage of such lower redshift observations is that these important feedback processes can be studied in a larger number of sources, greater detail and at higher resolution. 

We conclude that optical signatures of AGN feedback, such as the velocity of the AGN-driven wind, do not per se reflect the effect the AGN has on its host galaxy. Careful source selection has to be conducted in order to avoid possible biases. Specifically, we have shown in this work that investigating the relation between SFR and outflow strength leads to inconclusive results since wind-gas coupling strength corrections would have to be applied. Instead, if galaxies are selected based on their high gas content, i.e. based on their SFR, the coupling between wind and gas is strongest. A negative correlation between sSFR and outflow strength is observed which is driven by an increase in stellar mass of the host galaxies and is in agreement with a negative relative impact of AGN quenching their host galaxies.

\section*{Acknowledgements}
D.W. acknowledges support by the Akbari-Mack Postdoctoral Fellowship. 



\bibliographystyle{mnras}
\bibliography{master_bib}





\bsp	
\label{lastpage}
\end{document}